\documentclass[10pt, twocolumn]{extarticle}
\usepackage{ametsoc2col}
\usepackage{abstract}
\newcommand{\myabstract}{
\vspace{12pt}     This paper presents numerical solutions and idealized analytical 
solutions of axisymmetric, $f$-plane models of the tropical cyclone 
boundary layer. In the numerical model, the 
boundary layer radial and tangential flow is forced by a specified
pressure field, which can also be interpreted as a specified gradient 
balanced tangential wind field $v_{\rm gr}(r)$ or vorticity field 
$\zeta_{\rm gr}(r)$. When the specified 
$\zeta_{\rm gr}(r)$ field is changed from one that is radially 
concentrated in the inner core to one that is radially spread, the 
quasi-steady-state boundary layer flow transitions from a single eyewall 
shock-like structure to a double eyewall shock-like structure. To 
better understand these structures, analytical solutions are
presented for two simplified versions of the model. In the simplified
analytical models, which do not include horizontal diffusion, the
$u(\partial u/\partial r)$ term in the radial equation of motion and the
$u[f+(\partial v/\partial r)+(v/r)]$ term in the tangential equation 
of motion produce discontinuities in the radial and tangential wind, 
with associated singularities in the boundary layer pumping and the 
boundary layer vorticity. In the numerical model, which does include 
horizontal diffusion, the radial and tangential wind structures are 
not true discontinuities, but are shock-like, with wind changes of 
20 or 30 m~s$^{-1}$ over a radial distance of a few kilometers.  
When double shocks form, the outer shock can control the strength 
of the inner shock, an effect that likely plays an important role 
in concentric eyewall cycles. 
\vspace{12pt}}
\begin{document}
%
%%%%%%%%%%%%%%%%%%%%%%%%%%%%%%%%%%%%%%%%%%%%%%%%%%%%%%%%%%%%%%%%%%%%%
% TITLE
%
% Enter your TITLE here
%%%%%%%%%%%%%%%%%%%%%%%%%%%%%%%%%%%%%%%%%%%%%%%%%%%%%%%%%%%%%%%%%%%%%
\title{\textbf{\large{Tropical Cyclone Boundary Layer Shocks}}}
%
% Author names, with corresponding author information. 
% [Update and move the \thanks{...} block as appropriate.]
%
\author{\textsc{Christopher J.\ Slocum}
				\thanks{\textit{Corresponding author address:} 
				Christopher J.\ Slocum,\newline Department of Atmospheric Science,
                Colorado State University,\newline Fort Collins, Colorado, USA. 
				\newline{E-mail: cslocum@atmos.colostate.edu}}\\
\textit{\footnotesize{Department of Atmospheric Science,
Colorado State University, Fort Collins, Colorado, USA}}
\and 
\centerline{\textsc{Gabriel J. Williams}}\\% Add additional authors, different insitution
\centerline{\textit{\footnotesize{Department of Physics and Astronomy,
College of Charleston, Charleston, South Carolina, USA}}}
\and 
\centerline{\textsc{Richard K. Taft and Wayne H. Schubert}}\\% Add additional authors, different insitution
\centerline{\textit{\footnotesize{Department of Atmospheric Science,
Colorado State University, Fort Collins, Colorado, USA}}}
}
%
% Formatting done here...Authors should skip over this.  See above for abstract.
\ifthenelse{\boolean{dc}}
{
\twocolumn[
\begin{@twocolumnfalse}
\amstitle

% Start Abstract (Enter your Abstract above.  Do not enter any text here)
\begin{center}
\begin{minipage}{13.0cm}
\begin{abstract}
	\myabstract
	\newline
	\begin{center}
		\rule{38mm}{0.2mm}
	\end{center}
	\vspace{12pt}
\end{abstract}
\end{minipage}
\end{center}
\end{@twocolumnfalse}
]\saythanks
}
{
\amstitle
\begin{abstract}
\myabstract
\end{abstract}
\newpage
}
%%%%%%%%%%%%%%%%%%%%%%%%%%%%%%%%%%%%%%%%%%%%%%%%%%%%%%%%%%%%%%%%%%%%%
% MAIN BODY OF PAPER
%%%%%%%%%%%%%%%%%%%%%%%%%%%%%%%%%%%%%%%%%%%%%%%%%%%%%%%%%%%%%%%%%%%%%
\section{Introduction}
    
     \citet{williams13} interpreted the structure of the boundary layer 
wind field in Hurricane Hugo (1989) in terms of an axisymmetric slab boundary 
layer model. They explained Hugo's 20 m~s$^{-1}$ eyewall vertical velocity at 
450 m height by dry dynamics, i.e., by the formation of a shock-like structure 
in the boundary layer radial inflow, with small radial flow on the inside edge 
and large radial inflow on the outside edge of the structure. Such features 
appear to be primarily a phenomenon of the boundary layer because the radial
flow is an order of magnitude larger in the boundary layer compared to the
overlying fluid (typically 20 m~s$^{-1}$ versus 2 m~s$^{-1}$). Large inflow
in the boundary layer also leads to large tangential wind tendencies, resulting
in a shock-like structure in the boundary layer tangential wind. Since the radial
derivative of the radial velocity is related to the boundary layer pumping, and
the radial derivative of the tangential velocity is related to the vertical
component of relative vorticity, a thin annulus of very large boundary layer
pumping and very high vorticity develops as part of the shock structure.

     The purpose of the present paper is to add analytical support to the 
arguments of \citet{williams13} and to extend their work to the concentric 
eyewall case. The theoretical basis for the present arguments is again 
the axisymmetric, primitive equation version of the slab boundary layer model.  
% \citep{smith+montgomery08,smith+montgomery10,kepert10a,kepert10b,kepert13}  
The interpretation is again in terms of Burgers' shock effects, for which 
an excellent general mathematical discussion can be found in the book 
by \citet{whitham74}.

     The paper is organized in the following way. Section 2 gives a brief
review of the slab boundary layer model that was described in detail by
\citet{williams13}.  In order to gain a semi-quantitative understanding of
the solutions to the slab boundary layer equations, sections 3 and 4 present some
analytical solutions of simplified versions of the model. 
These solutions illustrate the formation of
discontinuities in the radial and tangential winds, and thus the formation of
singularities in the boundary layer pumping and the boundary layer vorticity. 
One application of these analytical solutions is to the formation of 
concentric eyewalls, which is discussed in section 5.
Numerical solutions of the complete nonlinear model are presented in section 6. 
In the numerical model, horizontal diffusion terms are used to maintain 
single-valued solutions, so that the numerically modeled structures are 
``shock-like" rather than true ``shocks", although for convenience we shall 
use these terms somewhat interchangeably. The numerical solutions 
are used to better understand the role of an outer eyewall shock in controlling
the structure of an inner eyewall shock. Section 7 presents some concluding
remarks, including the implications of the present work on understanding
eyewall replacement cycles. 
 
\section{Slab Boundary Layer Model}                     %%%%%%% Section 2 %%%%%%

    The model considers axisymmetric, boundary layer motions of an 
incompressible fluid on an $f$-plane.  The frictional boundary layer is assumed 
to have constant depth $h$, with radial and azimuthal velocities $u(r,t)$ and 
$v(r,t)$ that are independent of height between the top of a thin surface layer 
and height $h$, and with vertical velocity $w(r,t)$ at height $h$.  In
the overlying layer the radial velocity is assumed to be negligible and the
azimuthal velocity $v_{\rm gr}(r,t)$ is assumed to be in gradient balance 
and to be a specified function of radius and time. The boundary layer flow 
is driven by the same radial
pressure gradient force that occurs in the overlying fluid, so that, in the
radial equation of boundary layer motion, the pressure gradient force can be
expressed as the specified function $[f+(v_{\rm gr}/r)]v_{\rm gr}$. The governing
system of differential equations for the boundary layer variables $u(r,t)$,
$v(r,t)$, and $w(r,t)$ then takes the form
\begin{align}                                     % Equations (2.1 - 2.3)
      \frac{\partial u}{\partial t}
  &+ u\frac{\partial u}{\partial r} + w^- \left(\frac{u}{h}\right)
   = \left(f + \frac{v+v_{\rm gr}}{r}\right)\left(v - v_{\rm gr}\right)
  \notag \\
  & \quad - c_D U \frac{u}{h}
   +  K\frac{\partial}{\partial r}\left(\frac{\partial(ru)}{r\partial r}\right),
  \label{eq2.1} \\[1.5ex]
       \frac{\partial v}{\partial t}
  &+ u\left(f + \frac{\partial v}{\partial r} + \frac{v}{r}\right)
   + w^- \left(\frac{v - v_{\rm gr}}{h}\right)   \notag \\
  & \quad = -c_D U \frac{v}{h}
   + K\frac{\partial}{\partial r}\left(\frac{\partial(rv)}{r\partial r}\right),
  \label{eq2.2} \\[1.5ex]
     w &= -h\frac{\partial(ru)}{r\partial r}  \quad {\rm and} \quad  w^- = \tfrac{1}{2}(|w| - w),
  \label{eq2.3}
\end{align}
where
\begin{equation}                                 % Equation (2.4)
     U = 0.78\left(u^2 + v^2\right)^{1/2}
\label{eq2.4}
\end{equation}
is the wind speed at 10 m height, $f$ the constant Coriolis parameter, 
and $K$ the constant horizontal diffusivity. The drag coefficient $c_D$  
is assumed to depend on the 10 m wind speed according to 
\begin{equation}                                 % Equation (2.5)
     c_D = 10^{-3}
   \begin{cases}
      2.70/U + 0.142 + 0.0764 U
               \hspace*{0.2in}         {\rm if}\,\, U \le 25 \\[1.5ex]
      2.16 + 0.5406\left\{1-\exp[-(U-25)/7.5]\right\}        \\
               \hspace*{1.78in}        {\rm if}\,\, U \ge 25,
   \end{cases}
\label{eq2.5}
\end{equation}
where the 10 m wind speed $U$ is expressed in m~s$^{-1}$. The boundary 
conditions are 
\begin{equation}                                 % Equation (2.6)
     \left. \begin{array}{r} u = 0 \\ v = 0 \end{array} \right\}
         \;\; \mbox{at} \,\,\, r = 0,  \qquad 
     \left. \begin{array}{r} \dfrac{\partial(ru)}{\partial r} = 0 \\[1.5ex]
                             \dfrac{\partial(rv)}{\partial r} = 0
            \end{array} \right\}
         \;\; \mbox{at} \,\,\, r = b,
\label{eq2.6}
\end{equation}
where $b$ is the radius of the outer boundary. The initial conditions are 
\begin{equation}                                    % Equation (2.7)
         u(r,0) = u_0(r)  \quad {\rm and} \quad v(r,0) = v_0(r),  
\label{eq2.7}
\end{equation}
where $u_0(r)$ and $v_0(r)$ are specified functions. The forcing
$v_{\rm gr}(r,t)$ is discussed in section 6. 

     Applications of the slab boundary layer model (\ref{eq2.1})--(\ref{eq2.7}), 
or closely related models, have a rich history in the literature of hurricane 
dynamics. For at least a partial appreciation of this history, the reader is 
referred to the analyses and numerical simulations found in \citet{ooyama69a,ooyama69b}, 
\citet{chow71}, \citet{shapiro83}, \citet{emanuel86,emanuel89}, \citet{kepertwang01}, 
\citet{kepert01,kepert10a,kepert10b,kepert13},  
\citet{smith03}, \citet{smithvogl08}, \citet{smithetal08,smith09}, 
\citet{smith08,smith10a}, \citet{smith10b}, \citet{slocum13}, and \citet{abarca13}. 
As in the work of \citet{williams13}, our emphasis here is on high resolution 
simulations that capture the shock formation process. 
 
     In the absence of the horizontal diffusion terms, the slab boundary 
layer equations constitute a hyperbolic system that can be written 
in characteristic form (see the Appendix). A knowledge of the characteristic 
form is useful in understanding the formation of shocks. In fact, before presenting 
numerical solutions of the system (\ref{eq2.1})--(\ref{eq2.7}) in section 6, we next 
discuss some analytical solutions of two simplified versions of the model, i.e.,  
two versions that have very simple characteristic forms.  These
analytical solutions aid in understanding the formation of discontinuities 
in the radial and tangential flow, and hence singularities in the vertical 
velocity and vorticity.

\section{Analytical Model I}                          %%%%%%%% Section 3 %%%%%%%%

     The formation of shocks in the $u$ and $v$ fields in the hurricane 
boundary layer depends on the $u(\partial u/\partial r)$ and 
$u[f+(\partial v/\partial r)+(v/r)]$ terms in (\ref{eq2.1}) and (\ref{eq2.2}), 
with the term proportional to the agradient tangential flow $(v-v_{\rm gr})$ 
serving as a forcing mechanism for $(\partial u/\partial t)$, the surface friction 
terms serving to damp $u$ and $v$, and the horizontal diffusion terms serving 
to control the structure near the shock. As we shall see, the shocks in 
$u$ and $v$ occur at the same time and at the same radius. These discontinuities 
in the radial and tangential flow mean that there is a circle of 
infinite vertical velocity at the top of the boundary layer   
and a circular infinite vorticity sheet in the boundary layer. 
 
   To obtain a semi-quantitative understanding of the above concepts, 
we now approximate (\ref{eq2.1}) and (\ref{eq2.2})
by neglecting the horizontal diffusion terms, the $w^-$ terms, the surface 
drag terms, and the $(v - v_{\rm gr})$ forcing term.  
The radial and tangential momentum equations (\ref{eq2.1}) and
(\ref{eq2.2}) then simplify to  
\begin{equation}                                  % Equation (3.1)
     \frac{\partial u}{\partial t} + u\frac{\partial u}{\partial r} = 0, 
\label{eq3.1}
\end{equation}
\begin{equation}                                  % Equation (3.2)
                \frac{\partial v}{\partial t}
   + u\left(f + \frac{\partial v}{\partial r} + \frac{v}{r}\right) = 0.  
\label{eq3.2}
\end{equation}
Although the simplified equations (\ref{eq3.1}) and (\ref{eq3.2}) cannot 
be justified through a rigorous scale analysis of tropical cyclone boundary 
layer dynamics, they do contain an important part of the dynamics involved 
in boundary layer shocks. However, it 
should be noted that there is an important conceptual difference between 
the simple analytical model equations (\ref{eq3.1}) and (\ref{eq3.2}) 
and the original model equations (\ref{eq2.1}) and (\ref{eq2.2}). When we 
present numerical solutions of (\ref{eq2.1}) and (\ref{eq2.2}) in section 6, 
the initial condition will have no radial flow, so that a shock-like structure 
in $u$ will develop only after $u$ has been forced through the 
$[f+(v+v_{\rm gr})/r](v - v_{\rm gr})$ term in (\ref{eq2.1}). In contrast, 
the analytical solutions presented here will develop, not from 
this forcing effect, but rather from a nonzero initial condition on $u$.      

    The solutions of (\ref{eq3.1}) and (\ref{eq3.2}) are easily obtained 
by noting that these two equations can be written in the form  
\begin{equation}                                  % Equation (3.3)
      \frac{du}{dt} = 0,  
\label{eq3.3}
\end{equation} 
\begin{equation}                                  % Equation (3.4)
      \frac{d(rv+\frac{1}{2}fr^2)}{dt} = 0,  
\label{eq3.4}
\end{equation} 
where $(d/dt) = (\partial/\partial t) + u(\partial/\partial r)$ is the 
derivative following the boundary layer radial motion.  
According to (\ref{eq3.3}) and (\ref{eq3.4}), the radial velocity $u$ 
and the absolute angular momentum $rv+\frac{1}{2}fr^2$ are the Riemann 
invariants for analytical model I.  Integration of 
(\ref{eq3.3}) and (\ref{eq3.4}), with use of the initial conditions 
(\ref{eq2.7}), yields the solutions 
\begin{equation}                                  % Equation (3.5)
      u(r,t) = u_0(\hat{r}),  
\label{eq3.5}
\end{equation}
\begin{equation}                                  % Equation (3.6)
      v(r,t) = \Bigl(v_0(\hat{r}) + \tfrac{1}{2}f\hat{r}\Bigr) \frac{\hat{r}}{r}  
             - \tfrac{1}{2}fr,  
\label{eq3.6}
\end{equation} 
where the characteristics $\hat{r}(r,t)$ are given implicitly by 
\begin{equation}                                  % Equation (3.7)
     r = \hat{r} + tu_0(\hat{r}),   
\label{eq3.7}
\end{equation}
which is easily obtained by integration of $(dr/dt)=u$, with $u$ given 
by (\ref{eq3.5}).  For a given $\hat{r}$, (\ref{eq3.7}) defines a straight 
characteristic in $(r,t)$, along which the radial velocity $u(r,t)$ is 
fixed according to (\ref{eq3.5}), and along which the absolute angular 
momentum $rv(r,t)+\frac{1}{2}fr^2$ is fixed according to (\ref{eq3.6}).  

     To understand when the derivatives $(\partial u/\partial r)$ and 
$(\partial v/\partial r)$ become infinite, and to also check that 
(\ref{eq3.5}), (\ref{eq3.6}), and (\ref{eq3.7}) constitute solutions of 
(\ref{eq3.1}) and (\ref{eq3.2}), we first note that $(\partial/\partial t)$ and 
$(\partial/\partial r)$ of (\ref{eq3.7}) yield 
\begin{equation}                                  % Equation (3.8)
  \begin{split}
    -\frac{\partial\hat{r}}{\partial t} &= \frac{u_0(\hat{r})}{1+t u_0'(\hat{r})}, \\
     \frac{\partial\hat{r}}{\partial r} &= \frac{1           }{1+t u_0'(\hat{r})},
  \end{split}
\label{eq3.8}
\end{equation}
so that $(\partial/\partial t)$ and $u(\partial/\partial r)$ of (\ref{eq3.5}) yield 
\begin{equation}                                  % Equation (3.9)
  \begin{split}
        \frac{\partial u}{\partial t}  
    = u_0'(\hat{r})\frac{\partial\hat{r}}{\partial t}
   &= -\frac{u_0(\hat{r}) u_0'(\hat{r})}{1 + tu_0'(\hat{r})},  \\
      u\frac{\partial u}{\partial r} 
    = u_0(\hat{r})u_0'(\hat{r})\frac{\partial\hat{r}}{\partial r}
   &=  \frac{u_0(\hat{r}) u_0'(\hat{r})}{1 + tu_0'(\hat{r})}, 
  \end{split}
\label{eq3.9}
\end{equation}
where the final equalities in (\ref{eq3.9}) follow from using (\ref{eq3.8}) to 
eliminate $(\partial\hat{r}/\partial t)$ and $(\partial\hat{r}/\partial r)$.  
The sum of the two lines in (\ref{eq3.9}) then confirms that (\ref{eq3.5}) and 
(\ref{eq3.7}) constitute a solution of (\ref{eq3.1}). A similar argument
confirms that (\ref{eq3.6}) and (\ref{eq3.7}) constitute a solution of
(\ref{eq3.2}).  However, it should be noted that these solutions may be
multivalued, in which case (\ref{eq3.5})--(\ref{eq3.7}) must be amended in
such a way as to guarantee the solutions are single valued. In other words,
after the shock formation time $t_s$, a shock-tracking procedure is required.
To compute $t_s$ we note that, from the denominators on the right-hand sides
of (\ref{eq3.9}), the derivatives $(\partial u/\partial t)$ and
$(\partial u/\partial r)$ become infinite when    
\begin{equation}                                  % Equation (3.10)
       t u_0'(\hat{r}) = -1   
\label{eq3.10}
\end{equation}
along one or more of the characteristics. For a typical tropical cyclone, the
initial radial velocity profile $u_0(r)$ is such that its derivative
$u_0'(r)$ is both positive (for larger $r$) and negative (for smaller $r$).
Let $\hat{r}_s$ denote the characteristic that originates at the minimum
value of $u_0'(r)$, i.e., $u_0'(\hat{r}_s)=[u_0'(r)]_{\rm min}$.
Note that $u_0'(\hat{r}_s)$ will be the most negative value of $u_0'(\hat{r})$
and will satisfy (\ref{eq3.10}) at the earliest time.
Therefore, the time of shock formation, determined from (\ref{eq3.10}), is  
\begin{equation}                                 % Equation (3.11)
        t_s = -\frac{1}{u_0'(\hat{r}_s)},   
\label{eq3.11}
\end{equation}
and the radius of shock formation, determined from (\ref{eq3.7}) and 
(\ref{eq3.11}), is 
\begin{equation}                                 % Equation (3.12)
          r_s = \hat{r}_s - \frac{u_0(\hat{r}_s)}{u_0'(\hat{r}_s)}. 
\label{eq3.12}
\end{equation}
For typical tropical cyclone cases, $u_0(\hat{r}_s)<0$ and $u_0'(\hat{r}_s)<0$, 
so the shock forms a distance $u_0(\hat{r}_s)/u_0'(\hat{r}_s)$ inside 
$\hat{r}_s$. 

     From the solutions (\ref{eq3.5}) and (\ref{eq3.6}) we can compute the 
solutions for the divergence $\delta(r,t)=\partial[ru(r,t)]/r\partial r$ 
and the relative vorticity $\zeta(r,t)=\partial[rv(r,t)]/r\partial r$ .  
The relative vorticity is obtained by differentiation of (\ref{eq3.6}), 
which yields 
\begin{equation}                                  % Equation (3.13)
    \zeta(r,t) = \left(\frac{f + \zeta_0(\hat{r})}{1 + tu_0'(\hat{r})}\right)\frac{\hat{r}}{r} - f,  
\label{eq3.13}
\end{equation}
where $\zeta_0(r)=\partial[rv_0(r)]/r\partial r$
is the initial relative vorticity. Similarly, the boundary layer divergence  
$\delta(r,t)$, or equivalently the boundary layer pumping $w(r,t)=-h\delta(r,t)$, 
is obtained by using (\ref{eq3.5}) in (\ref{eq2.3}), which yields 
\begin{equation}                                  % Equation (3.14)
      w(r,t) = -h\left(\frac{u_0'(\hat{r})}{1 + tu_0'(\hat{r})}
		     + \frac{u_0 (\hat{r})}{r}\right).  
\label{eq3.14}
\end{equation}
Because of the factors $1 + tu_0'(\hat{r})$ in the denominators of (\ref{eq3.13}) 
and (\ref{eq3.14}), the relative vorticity $\zeta(r,t)$ and the boundary layer 
pumping $w(r,t)$ become infinite at the same time ($t=t_s$) and the same place ($r=r_s$).

    As a simple example, consider the initial conditions 
\begin{equation}                                  % Equation (3.15)
       u_0(r) = u_m \left(\frac{4(r/a)^3}{1 + 3(r/a)^4}\right),  
\label{eq3.15}
\end{equation}
\begin{equation}                                  % Equation (3.16)
       v_0(r) = v_m \left(\frac{2(r/a)}{1 + (r/a)^2}\right),  
\label{eq3.16}
\end{equation}
where the constants $a$, $u_m$, and $v_m$ specify the radial extent  
and strength of the initial radial and tangential flow.
The derivative of (\ref{eq3.15}) is 
\begin{equation}                                  % Equation (3.17)
       u_0'(r) = \frac{12u_m}{a}\left(\frac{(r/a)^2[1 - (r/a)^4]}{[1 + 3(r/a)^4]^2}\right),    
\label{eq3.17}
\end{equation}
while the initial relative vorticity, obtained by differentiation of (\ref{eq3.16}), is 
\begin{equation}                                  % Equation (3.18)
       \zeta_0(r) = \frac{4v_m}{a\left[1 + (r/a)^2\right]^2}.   
\label{eq3.18}
\end{equation} 

%%%%%%%%%%%%%%%%%%%%%%%%%%%%%%%%%%%%%%%%%%%%%%%%%%%%%%%%%%%%%%%%%%%%%%%%%%%%%%%
\begin{figure}[!t]              % Figure 1 (Initial Condition)
\centerline{\includegraphics[width=19pc]{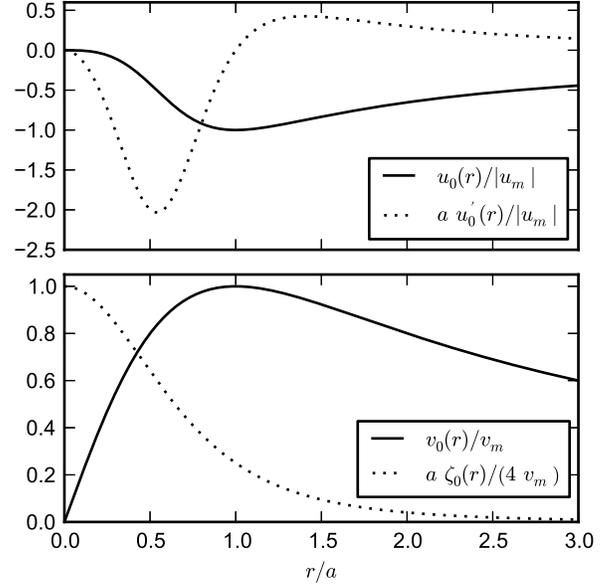}}
\caption{The dimensionless initial conditions used in the
analytical models for the single eyewall cases as computed from equations
(\ref{eq3.15})--(\ref{eq3.18}).  The solid line in the upper panel shows the
dimensionless initial radial velocity $u_0(r)/|u_m|$ for the case in which
$u_m<0$, while the dotted line shows its dimensionless radial derivative
$a \, u_0'(r)/|u_m|$.  Similarly, the solid line in the lower panel shows the
dimensionless initial tangential velocity $v_0(r)/v_m$, while the dotted line
shows the dimensionless initial vorticity $a \, \zeta_0(r)/(4v_m)$.}
\end{figure}
%%%%%%%%%%%%%%%%%%%%%%%%%%%%%%%%%%%%%%%%%%%%%%%%%%%%%%%%%%%%%%%%%%%%%%%%%%%%%%%
%%%%%%%%%%%%%%%%%%%%%%%%%%%%%%%%%%%%%%%%%%%%%%%%%%%%%%%%%%%%%%%%%%%%%%%%%%%%%%%
\begin{table*}[!t]                 % Table 1
\centering
\caption{Test Cases and Results for Analytical Models I and II}
\begin{center}
\begin{tabular}{*{13}{c}}
\hline \\[-1.5ex]
 & 
\multicolumn{3}{c}{Parameters Defining} &
\multicolumn{2}{c}{Typical Values} &
\multicolumn{4}{c}{Shock Results From} \\
 &
\multicolumn{3}{c}{Initial Conditions} &
\multicolumn{2}{c}{Needed in Model II} &
\multicolumn{4}{c}{Models I and II} \\[0.5ex]
\hline \\[-1.5ex]
Test & $a$  & $u_m$        & $v_m$        & $U$
        & $\tau$ & Shock & $r_s$        & $t^{(I)}_s$  & $t^{(II)}_s$ \\[0.5ex]
Case & (km) & (m~s$^{-1}$) & (m~s$^{-1}$) & (m~s$^{-1}$)
        & (h)    &       & (km)         & (h)          & (h) \\[1.0ex]
\hline \\[-1.5ex]
S1 & 300 & \phantom{0}$-0.5$ & \phantom{0}3.2 & \phantom{0}2.5 & 78.6\phantom{0} & Single & 87.9\phantom{0} & 82.0\phantom{00} & No Shock
  \\[0.5ex]
S2 & 200 & \phantom{0}$-1.0$ & \phantom{0}6.3 & \phantom{0}5.0 & 52.2\phantom{0} & Single & 58.6\phantom{0} & 27.3\phantom{00} & 38.7\phantom{00}
  \\[0.5ex]
S3 & 150 & \phantom{0}$-2.0$ & 12.7 & 10.0 & 23.6\phantom{0} & Single & 44.0\phantom{0} & 10.2\phantom{00} & 13.4\phantom{00}
  \\[0.5ex]
S4 & 100 & \phantom{0}$-4.0$ & 25.3 & 20.0 & \phantom{0}7.69 & Single & 29.3\phantom{0} & \phantom{0}3.42\phantom{0} & \phantom{0}4.52\phantom{0}
  \\[0.5ex]
S5 & \phantom{0}60 & \phantom{0}$-6.0$ & 38.0 & 30.0 & \phantom{0}3.82 & Single & 17.6\phantom{0} & \phantom{0}1.37\phantom{0}  & \phantom{0}1.69\phantom{0}
  \\[0.5ex]
S6 & \phantom{0}40 & \phantom{0}$-8.0$ & 50.7 & 40.0 & \phantom{0}2.64 & Single & 11.7\phantom{0} & \phantom{0}0.684 & \phantom{0}0.791
  \\[0.5ex]
S7 & \phantom{0}30 & $-10.0$ & 63.3 & 50.0 & \phantom{0}2.07 & Single & \phantom{0}8.79 & \phantom{0}0.410 & \phantom{0}0.457
  \\[0.5ex]
\hline \\[-1.5ex]
D1 & \phantom{0}60 & \phantom{0}$-6.0$ & 38.0 & 30.0 & \phantom{0}3.82 & Inner & 17.6\phantom{0} & \phantom{0}1.37\phantom{0} & \phantom{0}1.69\phantom{0}
  \\
   & & & & & & Outer & 29.6\phantom{0} & \phantom{0}2.43\phantom{0} & \phantom{0}3.85\phantom{0}
  \\[0.5ex]
\hline
\end{tabular}
\end{center}
\end{table*}
%%%%%%%%%%%%%%%%%%%%%%%%%%%%%%%%%%%%%%%%%%%%%%%%%%%%%%%%%%%%%%%%%%%%%%%%%%%%%%%

The dimensionless forms of the initial profiles (\ref{eq3.15})--(\ref{eq3.18}) 
are plotted in Figure 1. Note that $u_0'(r)=0$ at $r=0$ and $r=a$.  For this 
example, the minimum value of $u_0'(\hat{r})$ occurs at $\hat{r}=\hat{r}_s$, where 
\begin{equation}                                  % Equation (3.19)
       \hat{r}_s = \left(2 - \frac{\sqrt{33}}{3}\right)^{1/4} a \approx 0.5402 \, a, 
\label{eq3.19}
\end{equation}
so that, from (\ref{eq3.17}), 
\begin{equation}                                  % Equation (3.20)
       u_0'(\hat{r}_s) \approx 2.032 \frac{u_m}{a}.  
\label{eq3.20}
\end{equation}
From (\ref{eq3.11}), the time of shock formation is 
\begin{equation}                                  % Equation (3.21)
      t_s \approx -\frac{a}{2.032 \, u_m},  
\label{eq3.21}
\end{equation}
and, from (\ref{eq3.12}), the radius of shock formation is 
\begin{equation}                                  % Equation (3.22)
      r_s \approx 0.5426 \, \hat{r}_s \approx 0.2931 \, a.  
\label{eq3.22}
\end{equation}
Table 1 lists values of $t_s$ and $r_s$ obtained from model I for seven
different single eyewall test cases (S1 through S7) ranging from very weak
vortices to hurricane strength vortices.  For hurricane strength vortices,  
the shock formation time is generally less than 1 hour. These rapid shock 
formation times for strong vortices indicate that, if disrupted, hurricane 
eyewalls can rapidly reform. 

     For the initial conditions given by (\ref{eq3.15}) and (\ref{eq3.16}), 
the solutions (\ref{eq3.5}) and (\ref{eq3.6}) take the form    
\begin{equation}                                  % Equation (3.23)
      u(r,t) = u_m \left(\frac{4(\hat{r}/a)^3}{1 + 3(\hat{r}/a)^4}\right),  
\label{eq3.23}
\end{equation}
\begin{equation}                                  % Equation (3.24)
      v(r,t) = \left(\frac{2v_m(\hat{r}/a)}{1 + (\hat{r}/a)^2} 
                  + \tfrac{1}{2}f\hat{r}\right)\frac{\hat{r}}{r} 
              - \tfrac{1}{2} fr,  
\label{eq3.24}
\end{equation} 
where the characteristic curves (along which $\hat{r}$ is fixed) are defined by 
\begin{equation}                                  % Equation (3.25)
     r = \hat{r} + u_m t \left(\frac{4(\hat{r}/a)^3}{1+3(\hat{r}/a)^4}\right).   		                 
\label{eq3.25}
\end{equation}
Using (\ref{eq3.13}), the relative vorticity takes the form  
\begin{equation}                                  % Equation (3.26)
      \zeta(r,t) = \left(f+\frac{4v_m}{a\left[1 + (\hat{r}/a)^2\right]^2}\right)
                   \left(\frac{(\hat{r}/r)}{1 + tu_0'(\hat{r})}\right) - f,  
\label{eq3.26}
\end{equation}
while, using (\ref{eq3.14}), the boundary layer pumping takes the form 
\begin{equation}                                  % Equation (3.27)
  \begin{split}
   w(r,t) = &-\left(\frac{4hu_m (\hat{r}/a)^2}{a[1+3(\hat{r}/a)^4]}\right)  \\ 
            & \left(\frac{3[1-(\hat{r}/a)^4]}{[1 + tu_0'(\hat{r})][1+3(\hat{r}/a)^4]}
	          + \frac{\hat{r}}{r}\right).  
  \end{split}
\label{eq3.27}
\end{equation} 
The solutions for $u(r,t),\, v(r,t),\, \hat{r}(r,t)$, as given by 
(\ref{eq3.23})--(\ref{eq3.25}), are plotted in the two panels of Figure 2 for
the particular constants given in Table 1 for single eyewall test case S5
(i.e., $a=60$ km, $u_m=-6$ m~s$^{-1}$, $v_m=38$ m~s$^{-1}$).
The plots cover the radial interval $0 \le r \le 100$ km and the time interval
$0 \le t \le t_s$, where $t_s=1.37$ h is the shock formation time for this
particular initial condition. 
Another view of this analytical solution is given in Figure 3, with the four 
panels displaying the radial profiles (at $t=0$ in blue and at $t=t_s$ in red) 
of $u,v,w,\zeta$. Also shown by the black curves in the top two panels are
fluid particle displacements for particles that are equally spaced at the
initial time. 
At $t=t_s$ the $u$ and $v$ fields become discontinuous at $r=17.6$ km, while
the $w$ and $\zeta$ fields become singular there. 

%%%%%%%%%%%%%%%%%%%%%%%%%%%%%%%%%%%%%%%%%%%%%%%%%%%%%%%%%%%%%%%%%%%%%%%%%%%%%%%
\begin{figure*}[!t]             % Figure 2 (Single Eyewall Analytical Solution for Model I)
\centerline{\includegraphics[width=39pc]{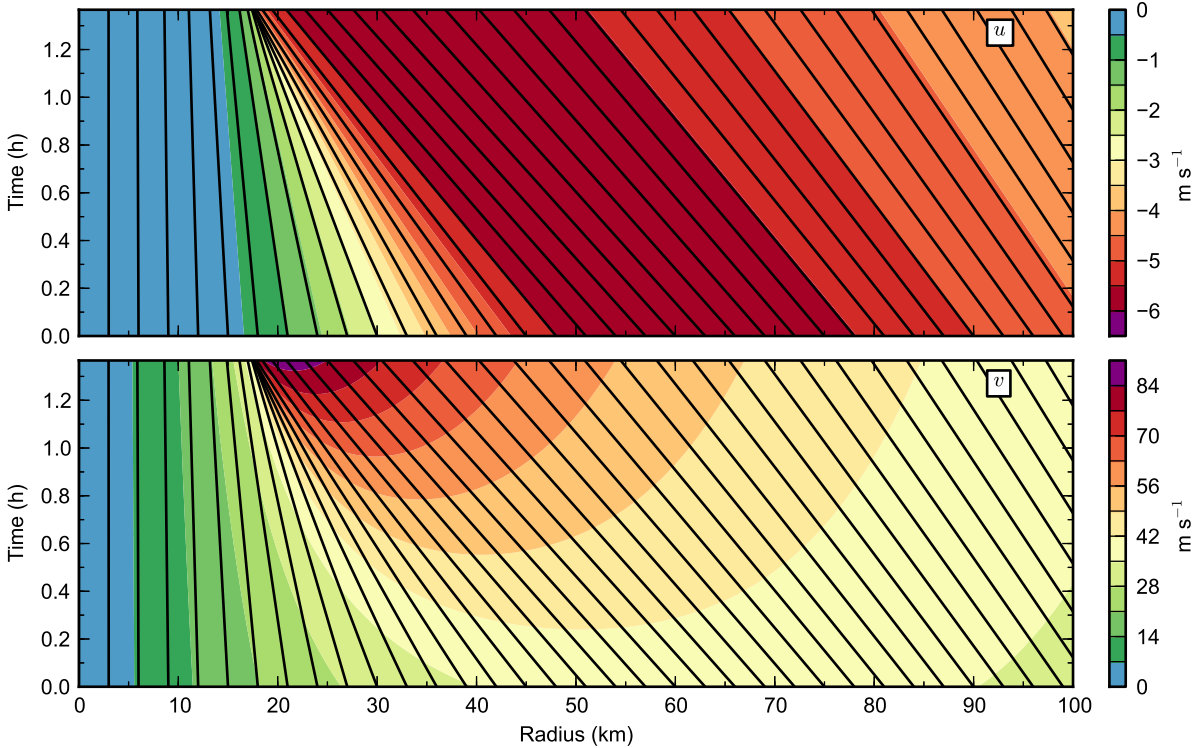}}
\caption{The analytical solutions $u(r,t)$ and $v(r,t)$ from model I
(color contours), as well as the characteristic curves (black lines
on each panel), for the single eyewall case.  These solutions are for the
particular initial conditions (\ref{eq3.15}) and (\ref{eq3.16}), with the
parameters given in Table 1 for test case S5
(i.e., $a=60$ km, $u_m=-6$ m~s$^{-1}$, $v_m=38$ m~s$^{-1}$).
The plots cover the time interval $0 \le t \le t_s$, where $t_s=1.37$ h is
the shock formation time for this particular initial condition.}
\end{figure*}

%%%%%%%%%%%%%%%%%%%%%%%%%%%%%%%%%%%%%%%%%%%%%%%%%%%%%%%%%%%%%%%%%%%%%%%%%%%%%%%

     The solutions plotted in Figure 2 cover the time interval
$0 \le t \le t_s$. How can we extend the solutions beyond $t=t_s$, i.e., 
into a region of the $(r,t)$-plane where characteristics intersect and
(\ref{eq3.23})--(\ref{eq3.25}) yield multivalued solutions?
One obvious way to address this issue is to return to the model equations 
(\ref{eq3.1})--(\ref{eq3.2}) and include horizontal diffusion terms. Indeed,
this is the approach that will control the shock-like structures in the
numerical solutions of section 6. However, even in the absence of horizontal
diffusion terms, we can amend the analytical solutions to guarantee they are
single-valued. One procedure is as follows.  
Let $R(t)$ denote the shock radius at time $t$, where $t \ge t_s$.
Let $\hat{r}_1(t)$ denote the label of the characteristic that just touches
the inside edge of the shock at time $t$, and $\hat{r}_2(t)$ denote the label
of the characteristic that just touches the outside edge of the shock at time
$t$. Then, from (\ref{eq3.7}), we obtain  
\begin{equation}                                  % Equation (3.28)
  \begin{split}
      R(t) &= \hat{r}_1(t) + t u_0(\hat{r}_1(t)),  \\   
      R(t) &= \hat{r}_2(t) + t u_0(\hat{r}_2(t)),  
  \end{split}
\label{eq3.28}
\end{equation}
which respectively determine $\hat{r}_1(t)$ and $\hat{r}_2(t)$ from a given
$R(t)$.  The last equation needed to track the shock is an appropriate jump
condition across the shock, which yields a first order ordinary differential 
equation relating $dR(t)/dt$ to $\hat{r}_1(t)$ and $\hat{r}_2(t)$.
Without going into the details of such arguments (see \citet{whitham74} for 
further discussion), we simply note that the solution of this ordinary 
differential equation for $R(t)$, along with (\ref{eq3.28}), 
yields the three functions $R(t)$, $\hat{r}_1(t)$, $\hat{r}_2(t)$.
This constitutes a shock-tracking procedure.
We shall not further explore this procedure, but rather simply note that such
shock-tracking procedures would probably never be used in three-dimensional, 
full-physics hurricane models because they become very complicated in more 
than one spatial dimension and when shocks can intersect. For a numerical model, 
a practical alternative to a shock-tracking procedure is a shock-capturing 
procedure, i.e., a procedure that captures the shock in a single grid volume 
for finite volume methods, or equivalently, between two grid points for 
finite difference methods. Shock capturing is a fundamental part 
of certain finite volume and finite difference methods based on 
the adaptive discretization concepts used in the essentially non-oscillatory 
(ENO) and the weighted essentially non-oscillatory (WENO) schemes (see 
the text by \citet{durran10} and the review by \citet{shu98}). Shock capturing 
is also part of the finite volume methods 
used in the software package CLAWPACK, which is described by \citet{leveque02}.  
With certain user-supplied routines, this software package
is capable of solving the nonlinear system (\ref{eq2.1})--(\ref{eq2.7}) without 
the horizontal diffusion terms. An advantage of using these shock-capturing 
methods is that they can reduce
both smearing and nonphysical oscillations near the discontinuity.
Although the numerical solutions of section 6 do not make use of shock  
capturing methods, these are interesting alternatives to the simple 
methods used here.  

    For the single eyewall case shown in Figures 2 and 3 the minimum value of
$u_0'(r)$ occurs at $r \approx 32.4$ km, and the shock forms at
$r \approx 17.6$ km.
In another class of initial conditions (not explored here) the minimum value
of $u_0'(r)$ occurs at $r=0$, in which case the shock forms at the center of
the vortex.  Although such initial conditions are probably less relevant,
they may explain certain features in simulated and real hurricanes, e.g.,
the very small ``vortex-within-a-vortex" sometimes seen in axisymmetric model
simulations (\citet{yamasaki83}, his Figures 18 and 20; \citet{hausman06},
their Figures 3 and 6) and the hub cloud sometimes seen in real hurricanes
\citep{schubert07}. 

%%%%%%%%%%%%%%%%%%%%%%%%%%%%%%%%%%%%%%%%%%%%%%%%%%%%%%%%%%%%%%%%%%%%%%%%%%%%%%%
\begin{figure}[!t]           % Figure 3 (Single Eyewall Analytical Solution for Model I)
\centerline{\includegraphics[width=19pc]{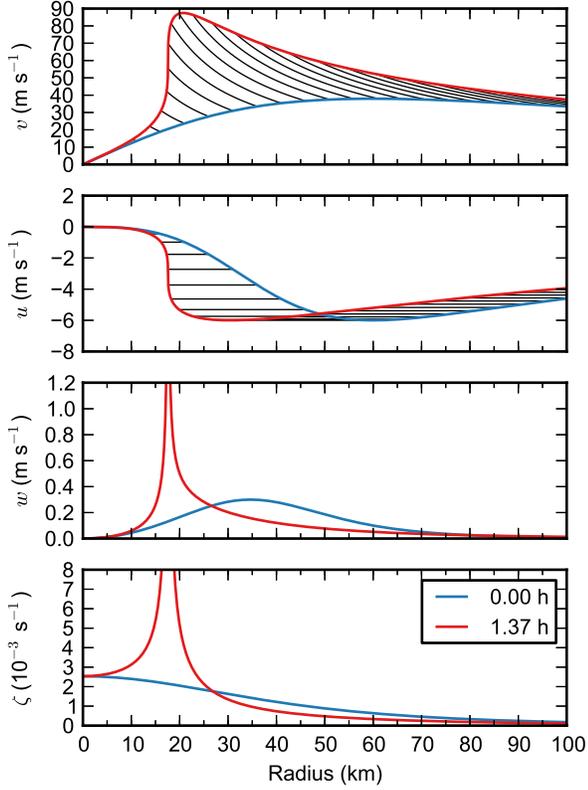}}
\caption{The radial profiles of $v,u,w,\zeta$ at $t=0$ (blue) and at
$t=t_s=1.37$ h (red) for analytical model I and single eyewall test case S5.
Also shown by the black curves in the top two panels are fluid particle
displacements for particles that are equally spaced at the initial time.
At $t=t_s$ the $u$ and $v$ fields become discontinuous at $r=r_s=17.6$ km,
while the $w$ and $\zeta$ fields become singular there.}
\end{figure}
%%%%%%%%%%%%%%%%%%%%%%%%%%%%%%%%%%%%%%%%%%%%%%%%%%%%%%%%%%%%%%%%%%%%%%%%%%%%%%%

%%%%%%%%%%%%%%%%%%%%%%%%%%%%%%%%%%%%%%%%%%%%%%%%%%%%%%%%%%%%%%%%%%%%%%%%%%%%%%%
\begin{figure}[!b]              % Figure 4 (Shock Formation Time)
\centerline{\includegraphics[width=19pc]{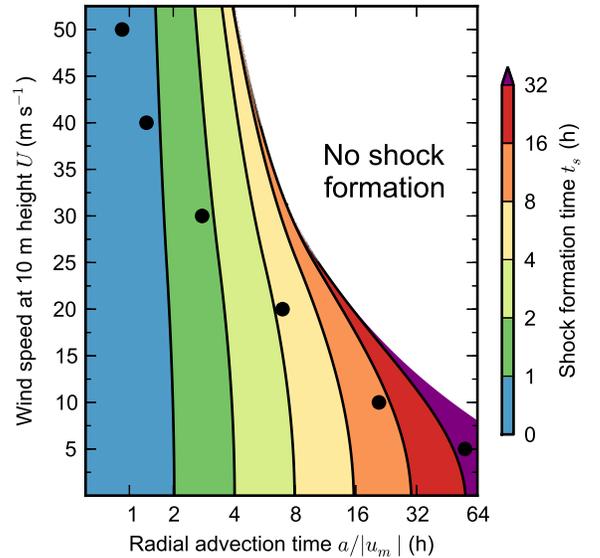}}
\caption{Isolines of the shock formation time $t_s$ (with color scale at right)
computed from (\ref{eq4.17}) as a function of the radial advection time
$a/|u_m|$ and the 10 m wind speed $U$. The six dots correspond to the
parameters given in Table 1 for the single eyewall test cases S2 through S7.}
\end{figure}
%%%%%%%%%%%%%%%%%%%%%%%%%%%%%%%%%%%%%%%%%%%%%%%%%%%%%%%%%%%%%%%%%%%%%%%%%%%%%%%
\section{Analytical Model II}                       %%%%%%%%  Section 4  %%%%%%%%%

     We now consider a second analytical model that adds surface drag effects 
to the model considered in section 3. For simplicity, we linearize the surface 
drag terms so the radial and tangential momentum equations become   
\begin{equation}                                  % Equation (4.1)
       \frac{\partial u}{\partial t}
    + u\frac{\partial u}{\partial r} = -\frac{u}{\tau}, 
\label{eq4.1}
\end{equation}
\begin{equation}                                  % Equation (4.2)
                \frac{\partial v}{\partial t}
   + u\left(f + \frac{\partial v}{\partial r} + \frac{v}{r}\right) = -\frac{v}{\tau}, 
\label{eq4.2}
\end{equation}
where the constant damping time scale $\tau$ is a typical value of $h/(c_D U)$.
The values of $\tau$ used for the test cases defined in Table 1
are given in the sixth column of that table and were computed using $h=1000$ m,
$c_D$ as given in (\ref{eq2.5}), and the values of $U$ given in the fifth
column of Table 1.  These typical values of $U$ were computed by finding the
maximum value of $U$ for the initial vortex of each test case.

    The solutions of (\ref{eq4.1}) and (\ref{eq4.2}) are easily obtained 
by noting that these two equations can be written in the form  
\begin{equation}                                  % Equation (4.3)
      \frac{d}{dt}\left\{ue^{t/\tau}\right\} = 0,  
\label{eq4.3}
\end{equation} 
\begin{equation}                                  % Equation (4.4)
      \frac{d}{dt}\left\{rve^{t/\tau} 
           + f\left[\hat{r}t + u_0(\hat{r})\tau(t-\hat{t})\right]u_0(\hat{r})\right\} = 0,  
\label{eq4.4}
\end{equation} 
where $(d/dt) = (\partial/\partial t) + u(\partial/\partial r)$ is again 
defined as the derivative following the boundary layer radial motion, and 
where the characteristics $\hat{r}(r,t)$ are given implicitly by 
\begin{equation}                                  % Equation (4.5)
     r = \hat{r} + \hat{t}u_0(\hat{r}),   
\label{eq4.5}
\end{equation}
with the function $\hat{t}(t)$ defined by 
\begin{equation}                                  % Equation (4.6)
     \hat{t} = \tau\left(1 - e^{-t/\tau}\right).    
\label{eq4.6}
\end{equation}
The quantities within the braces in (\ref{eq4.3}) and (\ref{eq4.4}) are 
the Riemann invariants for analytical model II. 
The equivalence of (\ref{eq4.2}) and (\ref{eq4.4}) is easily checked 
by converting (\ref{eq4.4}) to (\ref{eq4.2}) through the use of 
$\tau[d(t-\hat{t})/dt]=\hat{t}$, followed by the use of (\ref{eq4.5}). 

     Integration of (\ref{eq4.3}) and (\ref{eq4.4}), with use of the initial 
conditions (\ref{eq2.7}), yields the solutions 
\begin{equation}                                  % Equation (4.7)
      u(r,t) = u_0(\hat{r}) e^{-t/\tau},  
\label{eq4.7}
\end{equation}
\begin{equation}                                  % Equation (4.8)
   rv(r,t) = \Bigl\{\hat{r} v_0(\hat{r})    
           - f\left[ \hat{r}t + u_0(\hat{r})\tau(t - \hat{t})\right]u_0(\hat{r})\Bigr\} e^{-t/\tau}.   
\label{eq4.8}
\end{equation}
%%%%%%%%%%%%%%%%%%%%%%%%%%%%%%%%%%%%%%%%%%%%%%%%%%%%%%%%%%%%%%%%%%%%%%%%%%%%%%%
\begin{figure*}[!t]             % Figure 5 (Single Eyewall Analytic Solution for Model II)
\centerline{\includegraphics[width=39pc]{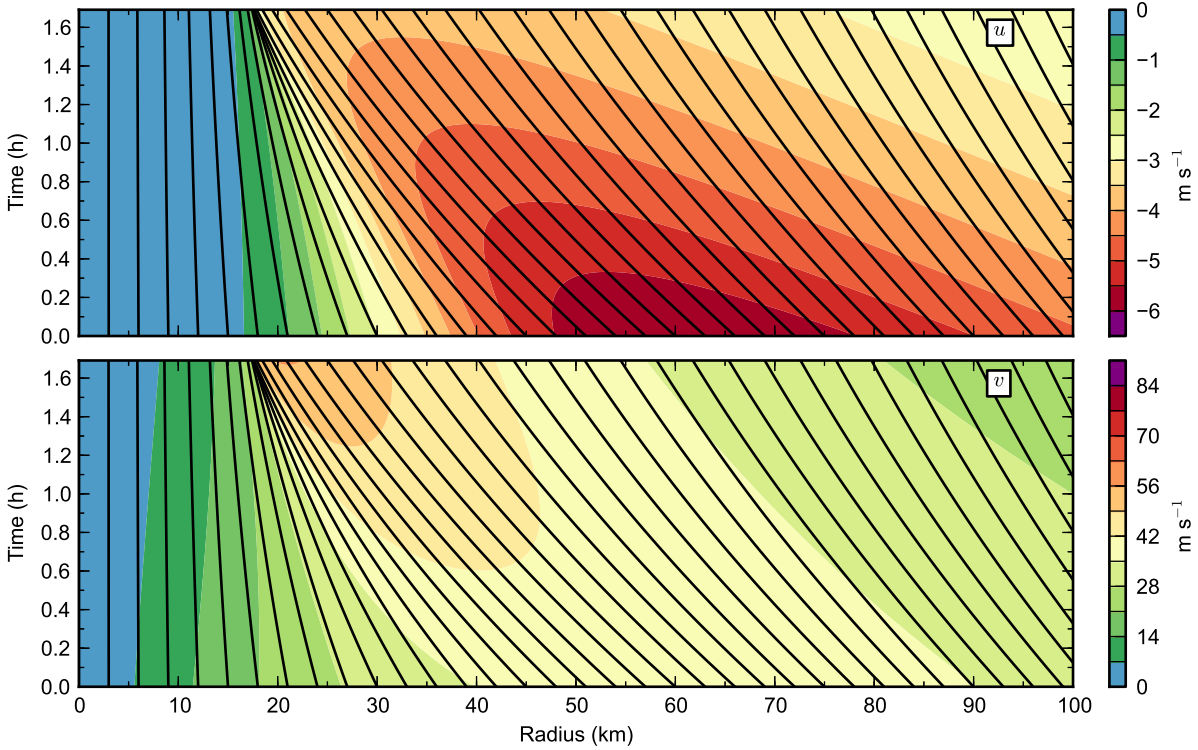}}
\caption{The analytical solutions $u(r,t)$ and $v(r,t)$ from model II
(color contours), as well as the characteristic curves (black curves
on each panel), for the single eyewall case. These solutions are for the
particular initial conditions (\ref{eq3.15}) and (\ref{eq3.16}), with the
parameters given in Table 1 for test case S5
(i.e., $a=60$ km, $u_m=-6$ m~s$^{-1}$, $v_m=38$ m~s$^{-1}$).
The plots cover the time interval $0 \le t \le t_s$, where $t_s=1.69$ h is
the shock formation time for model II with this particular initial condition.}
\end{figure*}
%%%%%%%%%%%%%%%%%%%%%%%%%%%%%%%%%%%%%%%%%%%%%%%%%%%%%%%%%%%%%%%%%%%%%%%%%%%%%%%
\noindent As in section 3, (\ref{eq4.5}) is easily obtained by integration of 
$(dr/dt)=u$, but with $u$ now given by (\ref{eq4.7}).  
For a given $\hat{r}$, (\ref{eq4.5}) defines a curved characteristic 
in $(r,t)$, along which $u(r,t)$ exponentially damps according 
to (\ref{eq4.7}), and along which, according to (\ref{eq4.8}), 
$v(r,t)$ varies in a more complicated way that includes the factor 
$(\hat{r}/r)e^{-t/\tau}$. Along a given characteristic, the behavior of $u(r,t)$
can be quite different from the behavior of $v(r,t)$ because the effect of the $(\hat{r}/r)$ 
amplification factor can more than compensate the $e^{-t/\tau}$ damping factor 
and cause $v(r,t)$ to increase along some characteristics. 

     Another useful representation of the solution 
(\ref{eq4.8}) is obtained by using (\ref{eq4.5}) to eliminate $u_0(\hat{r})$, 
yielding the form 
\begin{equation}                                  % Equation (4.9)
   \begin{split}
   rv(r,t) &= \biggl\{\hat{r} v_0(\hat{r})    \\  
           &+ f\frac{t}{\hat{t}}\left[ \hat{r} 
	    + (r-\hat{r})\left(\frac{\tau(t-\hat{t})}{t\hat{t}}\right)\right](\hat{r}-r)\biggr\} e^{-t/\tau},      		    
   \end{split}
\label{eq4.9}
\end{equation} 
%
%\begin{equation}                                  % Equation (4.9)
%   rv(r,t) = \biggl\{\hat{r} v_0(\hat{r})     
%           + f\frac{t}{\hat{t}}\left[ \hat{r} 
%	    + (r-\hat{r})\left(\frac{\tau(t-\hat{t})}{t\hat{t}}\right)\right](\hat{r}-r)\biggr\} e^{-t/\tau},      		    
%\label{eq4.9}
%\end{equation} 
%
which is analogous to the frictionless form (\ref{eq3.6}). 
In fact, for $(t/\tau) \ll 1$, it can be shown that $(t/\hat{t}) \approx 1$ and 
$\tau(t-\hat{t})/(t\hat{t}) \approx 1/2$, in which case (\ref{eq4.9}) reduces 
to (\ref{eq3.6}). 

     To understand when the derivatives $(\partial u/\partial r)$ and 
$(\partial v/\partial r)$ become infinite, we first note that $(\partial/\partial t)$ 
and $(\partial/\partial r)$ of (\ref{eq4.5}) yield 
\begin{equation}                                  % Equation (4.10)
  \begin{split}
    -\frac{\partial\hat{r}}{\partial t} &= \frac{u_0(\hat{r})e^{-t/\tau}}{1+\hat{t} u_0'(\hat{r})}, \\
     \frac{\partial\hat{r}}{\partial r} &= \frac{1                      }{1+\hat{t} u_0'(\hat{r})},
  \end{split}
\label{eq4.10}
\end{equation}
so that $(\partial/\partial t + 1/\tau)$ and $u(\partial/\partial r)$ of (\ref{eq4.7}) yield 
\begin{equation}                                  % Equation (4.11)
  \begin{split}
        \frac{\partial u}{\partial t} + \frac{u}{\tau}  
    = e^{-t/\tau} u_0'(\hat{r})\frac{\partial\hat{r}}{\partial t}
   &= -\frac{e^{-2t/\tau} \, u_0(\hat{r}) u_0'(\hat{r})}{1 + \hat{t} u_0'(\hat{r})},  \\
      u\frac{\partial u}{\partial r} 
    = e^{-2t/\tau} \, u_0(\hat{r})u_0'(\hat{r})\frac{\partial\hat{r}}{\partial r}
   &=  \frac{e^{-2t/\tau} u_0(\hat{r}) u_0'(\hat{r})}{1 + \hat{t} u_0'(\hat{r})}, 
  \end{split}
\label{eq4.11}
\end{equation}
where the final equalities in (\ref{eq4.11}) follow from using (\ref{eq4.10}) to 
eliminate $(\partial\hat{r}/\partial t)$ and $(\partial\hat{r}/\partial r)$.  
The sum of the two lines in (\ref{eq4.11}) then confirms that (\ref{eq4.5}) and 
(\ref{eq4.7}) constitute a solution of (\ref{eq4.1}). A similar argument
confirms that (\ref{eq4.5}) and (\ref{eq4.8}) constitute a solution of
(\ref{eq4.2}).  To compute $t_s$ we note that, from the denominators on the 
right-hand sides of (\ref{eq4.11}), the derivatives $(\partial u/\partial t)$ 
and $(\partial u/\partial r)$ can become infinite if   
\begin{equation}                                  % Equation (4.12)
       \hat{t} u_0'(\hat{r}) = -1   
\label{eq4.12}
\end{equation}
along one or more of the characteristics. The condition (\ref{eq4.12}) is 
identical to (\ref{eq3.10}), except that $\hat{t}$ has replaced $t$. This 
is an important difference because, by inspection of (\ref{eq4.6}), we 
note that $0 \le \hat{t} < \tau$ while $0 \le t < \infty$, i.e., the value of 
$\hat{t}$ may never get large enough to satisfy (\ref{eq4.12}), in which 
case a shock will not form. Shock formation is possible if and only if 
$\tau [u_0'(\hat{r}_s)] < -1$, where $\hat{r}_s$ again denotes the 
characteristic that originates at the minimum value of $u_0'(r)$, 
i.e., $u_0'(\hat{r}_s)=[u_0'(r)]_{\rm min}$.
In other words, if the initial radial velocity $u_0(r)$ has a large enough
negative slope, the solution will become multivalued.
Then, the time of shock formation, determined by combining (\ref{eq4.6}) 
and (\ref{eq4.12}), is  
\begin{equation}                                 % Equation (4.13)
        t_s = -\tau \ln\left(1 + \frac{1}{\tau u_0'(\hat{r}_s)}\right),   
\label{eq4.13}
\end{equation}
and the radius of shock formation, determined from (\ref{eq4.5}) and 
(\ref{eq4.12}), is 
\begin{equation}                                 % Equation (4.14)
          r_s = \hat{r}_s - \frac{u_0(\hat{r}_s)}{u_0'(\hat{r}_s)}. 
\label{eq4.14}
\end{equation}
Note that $t_s$ depends on the damping time scale $\tau$, but $r_s$ 
is independent of $\tau$ and identical to that found in model I. 

     From the solutions for $u(r,t)$ and $v(r,t)$ we can compute the 
solutions for the relative vorticity $\zeta(r,t)=\partial[rv(r,t)]/r\partial r$ 
and the divergence $\delta(r,t)=\partial[ru(r,t)]/r\partial r$.  The relative 
vorticity is obtained by differentiation of (\ref{eq4.8}), which yields 
\begin{equation}                                  % Equation (4.15)
  \begin{split}
    \zeta(r,t) &= \Biggl\{\left(\frac{\frac{t}{\hat{t}}\left[1 + \left(1 - \frac{r}{\hat{r}}\right)
                              \left(1 - \frac{2\tau(t-\hat{t})}{t\hat{t}}\right)\right]f 
			+ \zeta_0(\hat{r})} 
                       {1 + \hat{t}u_0'(\hat{r})}\right) \frac{\hat{r}}{r}     \\
  	       &- f\frac{t}{\hat{t}}\left[1 - \left(1 - \frac{\hat{r}}{r}\right)
		   \left(1 - \frac{2\tau(t-\hat{t})}{t\hat{t}}\right)\right]\Biggr\} e^{-t/\tau},  
  \end{split}
\label{eq4.15}
\end{equation}
where $\zeta_0(r)=\partial[rv_0(r)]/r\partial r$
is the initial relative vorticity.  Note that (\ref{eq4.15}) reduces to 
(\ref{eq3.13}) when $(t/\tau) \ll 1$. Similarly, the boundary layer divergence, or
equivalently the boundary layer pumping $w(r,t)=-h\delta(r,t)$, is obtained
by using (\ref{eq4.7}) in (\ref{eq2.3}), which yields 
\begin{equation}                                  % Equation (4.16)
      w(r,t) = -h\left(\frac{u_0'(\hat{r})}{1+\hat{t}u_0'(\hat{r})}
	             + \frac{u_0 (\hat{r})}{r}\right) e^{-t/\tau}.  
\label{eq4.16}
\end{equation}
Because of the presence of the $1+\hat{t}u_0'(\hat{r})$ term in the denominators 
of both (\ref{eq4.15}) and (\ref{eq4.16}), the relative vorticity $\zeta(r,t)$ 
and the boundary layer pumping $w(r,t)$ become infinite at the same time and place.

     As a simple example, again consider the initial conditions given 
in (\ref{eq3.15}) and (\ref{eq3.16}).  
From (\ref{eq4.13}) and (\ref{eq4.14}), the time and radius of shock
formation are 
\begin{equation}                                  % Equation (4.17)
  \begin{split}
      & t_s \approx -\tau\ln\left(1 - \frac{a}{2.032\tau |u_m|}\right), \\
      & r_s \approx 0.5426 \, \hat{r}_s \approx 0.2931 \, a.  
  \end{split}
\label{eq4.17}
\end{equation}
%%%%%%%%%%%%%%%%%%%%%%%%%%%%%%%%%%%%%%%%%%%%%%%%%%%%%%%%%%%%%%%%%%%%%%%%%%%%%%%

\begin{figure}[!t]             % Figure 6 (Single Eyewall Analytic Solution for Model II)
\centerline{\includegraphics[width=19pc]{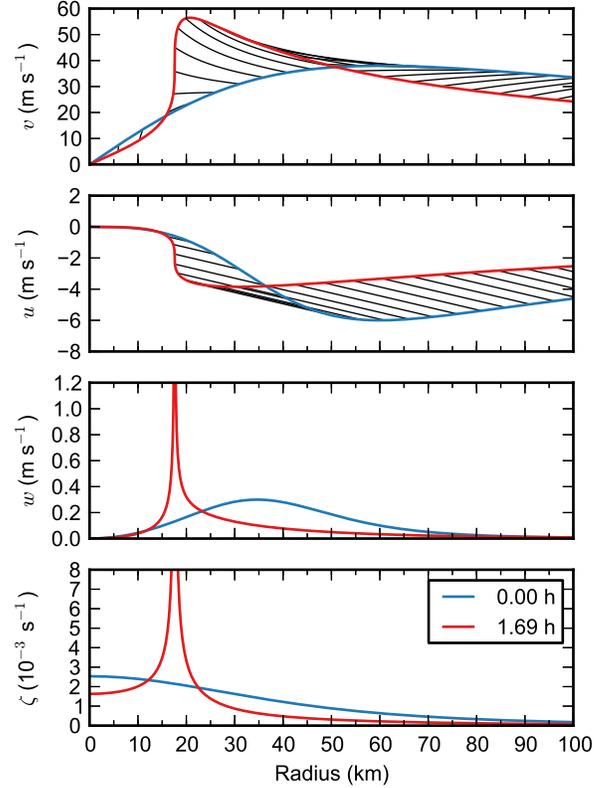}}
\caption{The radial profiles of $v,u,w,\zeta$ at $t=0$ (blue) and at
$t=t_s=1.69$ h (red) for analytical model II and single eyewall test case S5.
Also shown by the black curves in the top two panels are fluid particle
displacements for particles that are equally spaced at the initial time.
At $t=t_s$ the $u$ and $v$ fields become discontinuous at $r=r_s=17.6$ km,
while the $w$ and $\zeta$ fields become singular there.}
\end{figure}
\begin{figure}[!t]              % Figure 7 (Ratio of Shock Formation Times)
\centerline{\includegraphics[width=19pc]{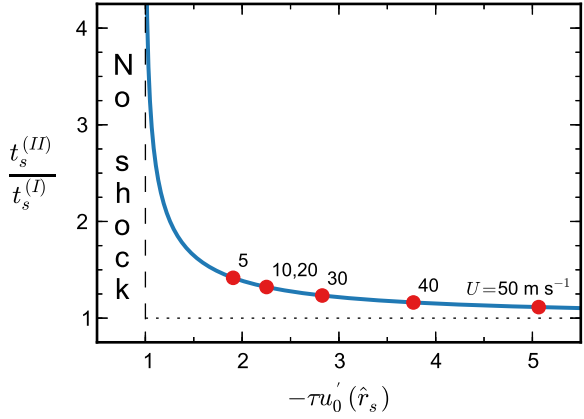}}
\caption{The ratio of shock formation times from the two analytical models as
computed from (\ref{eq4.23}). The red dots correspond to the values of $U$
given in Table 1 for the single eyewall test cases S2 through S7. Note that the
dots for $U=10$ m~s$^{-1}$ and for $U=20$ m~s$^{-1}$ are practically colocated.}
\end{figure}
%%%%%%%%%%%%%%%%%%%%%%%%%%%%%%%%%%%%%%%%%%%%%%%%%%%%%%%%%%%%%%%%%%%%%%%%%%%%%%%
\noindent The last column of Table 1 lists values of $t_s$ obtained from model II for
the seven different single eyewall test cases defined in that table.
Except for the weakest vortex case (S1), all cases produce boundary layer
shocks, i.e., the surface drag effects cannot prevent the development of
discontinuities in the $u$ and $v$ fields. For the hurricane strength vortices 
given in Table 1, the shock formation time is generally less than 1 hour.  
Since $\tau=h/(c_D U)$, the top line in (\ref{eq4.17}) can also be 
regarded as giving the shock formation time $t_s$ as a function of the radial
advection time $a/|u_m|$ and the 10 m wind speed $U$. Contours of $t_s$, as 
a function of $a/|u_m|$ and $U$ are shown in Figure 4. The six dots 
correspond to the parameters for single eyewall test cases S2 through S7 as
given in Table 1.

     For this initial condition, the solutions (\ref{eq4.7}) and (\ref{eq4.9}) 
take the form    
\begin{equation}                                  % Equation (4.18)
      u(r,t) = u_m \left(\frac{4(\hat{r}/a)^3 e^{-t/\tau}}{1 + 3(\hat{r}/a)^4}\right),  
\label{eq4.18}
\end{equation} 
\begin{equation}                                  % Equation (4.19)
   \begin{split}
   rv(r,t) &= \biggl\{\hat{r} v_m \left(\frac{2(\hat{r}/a)}{1 + (\hat{r}/a)^2}\right) \\  
           &+ f\frac{t}{\hat{t}}\left[ \hat{r} 
	    + (r-\hat{r})\left(\frac{\tau(t-\hat{t})}{t\hat{t}}\right)\right](\hat{r}-r)\biggr\} e^{-t/\tau},      		    
   \end{split}
\label{eq4.19}
\end{equation} 
where the characteristic curves (along which $\hat{r}$ is fixed) are defined by 
\begin{equation}                                  % Equation (4.20)
     r = \hat{r} + u_m \hat{t}  
                  \left(\frac{4(\hat{r}/a)^3}{1+3(\hat{r}/a)^4}\right).   		                 
\label{eq4.20}
\end{equation}
%%%%%%%%%%%%%%%%%%%%%%%%%%%%%%%%%%%%%%%%%%%%%%%%%%%%%%%%%%%%%%%%%%%%%%%%%%%%%%%
\begin{figure}[!b]              % Figure 8 (Initial Condition for Double Eyewall)
\centerline{\includegraphics[width=19pc]{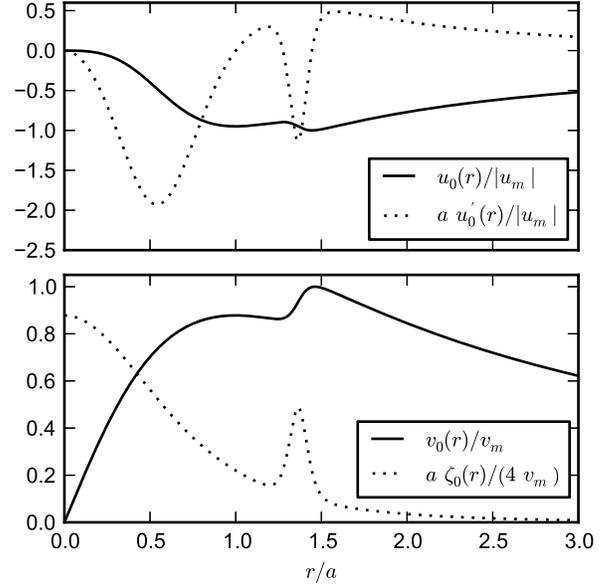}}
\caption{The dimensionless initial conditions used in
analytical model II for the double eyewall case. The solid line in the upper
panel shows the dimensionless initial radial velocity $u_0(r)/|u_m|$, while
the dotted line shows its dimensionless radial derivative $a \, u_0'(r)/|u_m|$.
Similarly, the solid line in the lower panel shows the dimensionless initial
tangential velocity $v_0(r)/v_m$, while the dotted line shows the dimensionless
initial vorticity $a \, \zeta_0(r)/(4v_m)$.}
\end{figure}
%%%%%%%%%%%%%%%%%%%%%%%%%%%%%%%%%%%%%%%%%%%%%%%%%%%%%%%%%%%%%%%%%%%%%%%%%%%%%%%
\noindent The formula for the relative vorticity can be obtained by using (\ref{eq3.18}) 
in (\ref{eq4.15}), while the formula for the boundary layer pumping can 
be obtained by using (\ref{eq3.15}) and (\ref{eq3.17}) in (\ref{eq4.16}). 
The solutions for $u(r,t),\, v(r,t),\, \hat{r}(r,t)$, as given by 
(\ref{eq4.18})--(\ref{eq4.20}), are plotted in the two panels of Figure 5 for
the particular initial parameters given in Table 1 for the single eyewall
test case S5 (i.e., $a=60$ km, $u_m=-6$ m~s$^{-1}$, $v_m=38$ m~s$^{-1}$).
The plots cover the radial interval $0 \le r \le 100$ km and the time interval
$0 \le t \le t_s$, where $t_s=1.69$ h is the shock formation time for this
particular initial condition. 
Another view of this analytical solution is given in Figure 6, with the four 
panels displaying the radial profiles (at $t=0$ in blue and at $t=t_s$ in red) 
of $u,v,w,\zeta$. Also shown by the black curves in the top two panels are
fluid particle displacements for particles that are equally spaced at the
initial time. 
At $t=t_s$ the $u$ and $v$ fields become discontinuous at $r=17.6$ km, while the $w$ and
$\zeta$ fields become singular there. 

     The shock formation times for models I and II are given by (\ref{eq3.11}) and 
(\ref{eq4.13}).  To see how the inclusion of surface friction lengthens the shock 
formation time, we can take the ratio of these two formulas to obtain 
\begin{equation}                                  % Equation (4.23)
   \frac{t_s^{(II)}}{t_s^{(I)}} = \tau u_0'(\hat{r}_s) 
                              \ln\left(1 + \frac{1}{\tau u_0'(\hat{r}_s)}\right),  
\label{eq4.23}
\end{equation}
where the superscripts I and II have been introduced to distinguish the two models. 
A graph of $t_s^{(II)}/t_s^{(I)}$ as a function of $-\tau u_0'(\hat{r}_s)$ is shown in Figure 7. 
As noted previously, shocks do not form in model II for $-\tau u_0'(\hat{r}_s) \le 1$.  
For $1 < -\tau u_0'(\hat{r}_s) < 3$ there are important differences in shock formation 
times from the two models, with the shock formation time lengthened by 38.6\% when  
$-\tau u_0'(\hat{r}_s)=2$ and by 21.6\% when $-\tau u_0'(\hat{r}_s)=3$.  
When $-\tau u_0'(\hat{r}_s) > 4$, the effects of surface friction do not substantially lengthen 
the shock formation time. For the example shown in Figures 5 and 6, 
$-\tau u_0'(\hat{r}_s)=2.8$, so that the shock formation time is lengthened by 
23.4\% due to surface friction effects.
%%%%%%%%%%%%%%%%%%%%%%%%%%%%%%%%%%%%%%%%%%%%%%%%%%%%%%%%%%%%%%%%%%%%%%%%%%%%%%%
\begin{figure*}[!t]              % Figure 9 (Double Eyewall Analytic Solution I)
\centerline{\includegraphics[width=39pc]{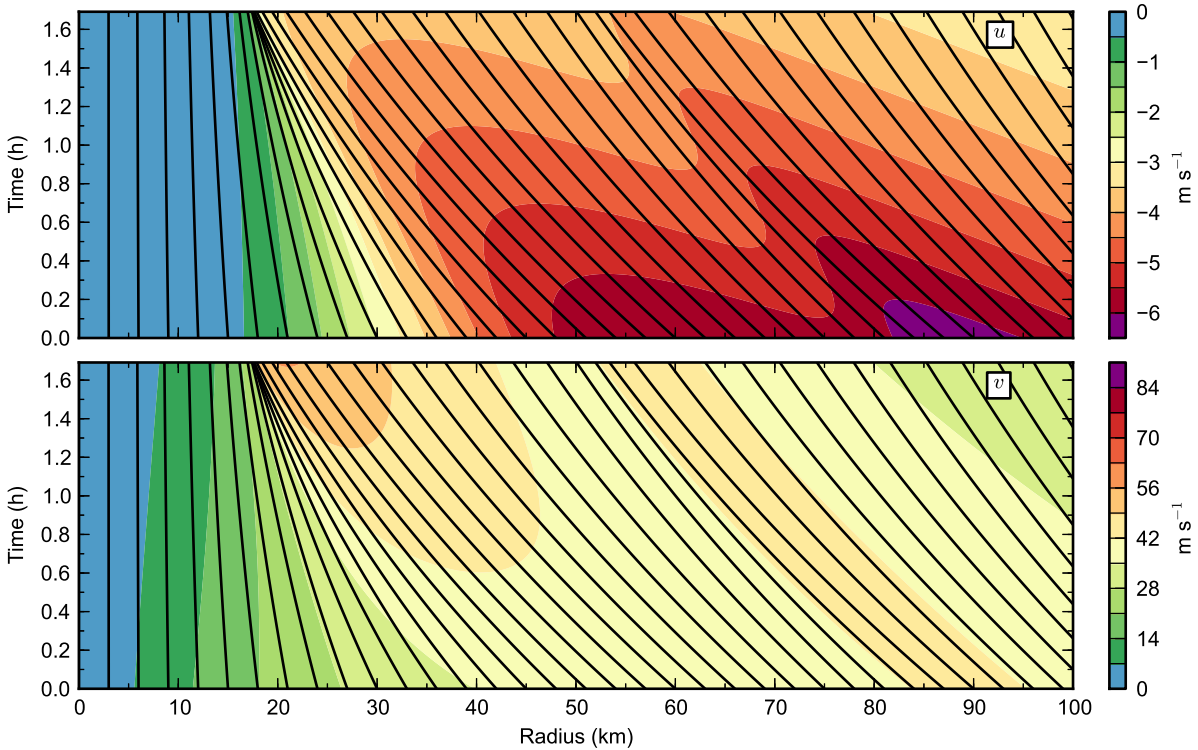}}
\caption{The analytical solutions $u(r,t)$ and $v(r,t)$ from model II
(color contours), as well as the characteristic curves (black curves
on each panel), for the double eyewall case. These solutions are for the
particular initial conditions shown in Figure 8, with the
parameters given in Table 1 for the double eyewall test case D1
(i.e., $a=60$ km, $u_m=-6$ m~s$^{-1}$, $v_m=38$ m~s$^{-1}$).
The plots cover the time interval $0 \le t \le t_s$, where $t_s=1.69$ h is
the shock formation time for model II with this particular initial condition.}
\end{figure*}
%%%%%%%%%%%%%%%%%%%%%%%%%%%%%%%%%%%%%%%%%%%%%%%%%%%%%%%%%%%%%%%%%%%%%%%%%%%%%%%
\noindent However, it is important to note that small 
differences in shock formation time do not imply small differences in the structures of 
the shocks. This can be seen by comparing the red curves in the upper two panels of Figure 3 
with the corresponding red curves in the upper two panels of Figure 6. For example, since 
the maximum tangential wind in Figure 6 is approximately 33 m~s$^{-1}$ weaker than the maximum 
tangential wind in Figure 3, surface friction has played an important role in reducing 
the angular momentum of the inflowing boundary layer air in Figure 6.  

%%%%%%%%%%%%%%%%%%%%%%%%%%%%%%%%%%%%%%%%%%%%%%%%%%%%%%%%%%%%%%%%%%%%%%%%%%%%%%%
\begin{figure}[!b]             % Figure 10 (Double Eyewall Analytic Solution II)
\centerline{\includegraphics[width=19pc]{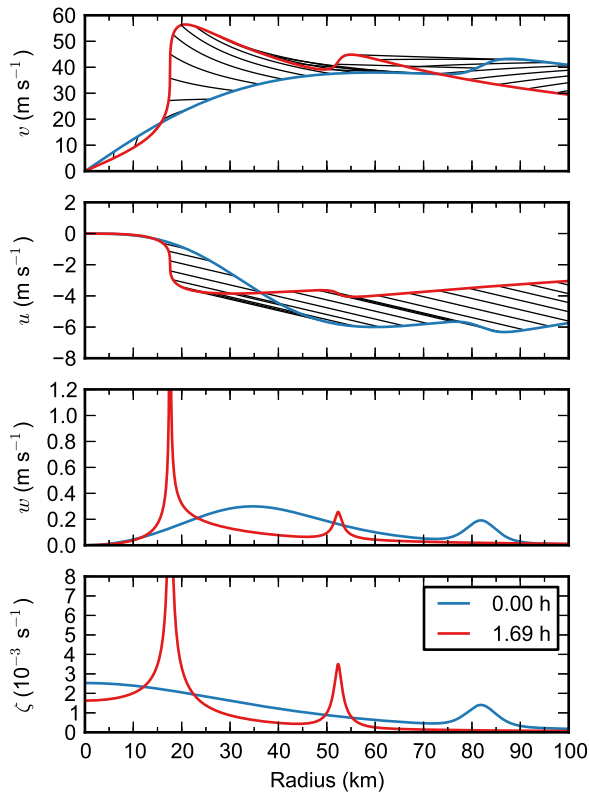}}
\caption{The radial profiles of $u,v,w,\zeta$ at $t=0$ (blue) and at
$t=t_s=1.69$ h (red) for analytical model II and double eyewall test case D1.
Also shown by the black curves in the top two panels are fluid particle
displacements for particles that are equally spaced at the initial time.
At $t=t_s$ the $u$ and $v$ fields become discontinuous at $r=17.6$ km, while
the $w$ and $\zeta$ fields become singular there.}
\end{figure}
%%%%%%%%%%%%%%%%%%%%%%%%%%%%%%%%%%%%%%%%%%%%%%%%%%%%%%%%%%%%%%%%%%%%%%%%%%%%%%%
\section{Double Eyewall Case}           %%%%%%%%%  Section 5  %%%%%%%%%

     In recent years there has been remarkable progress in our understanding 
of secondary eyewall formation and concentric eyewall cycles. This progress 
includes aircraft-based and satellite-based observational analyses (e.g., \citet{willoughby82}, 
\citet{black92}, \citet{samsury95}, \citet{dodge99}, \citet{bell12}, 
\citet{hence12}, \citet{yang13}), operational detection and forecasting 
(e.g., \citet{willoughby96}, \citet{maclay08}, \citet{kossin09,kossin12}, 
\citet{sitkowski11,sitkowski12}), analytical analyses and numerical simulations 
using idealized models (e.g., \citet{shapiro82}, \citet{nong03}, \citet{kuo04,kuo08,kuo09}, 
\citet{rozoff06,rozoff08}, \citet{menelaou13}), 
and, most recently, numerical simulations with three-dimensional full-physics models 
(e.g., \citet{houze07,terwey08,wang08,wang09,zhou09,judt10,abarca11,martinez11,rozoff12,wu12,  
huang12,menelaou12,lee12,chen13,wang13,abarca13}). These three-dimensional 
simulations, although run at coarser horizontal resolutions than the
present axisymmetric slab model, can be interpreted as demonstrating 
the importance of the boundary layer shock phenomenon. 

     In order to better understand the formation of concentric eyewalls, we 
now consider solutions of analytical model II for an initial condition that 
leads to double shocks. In this example, the initial dimensionless radial 
wind $u_0(r)/|u_m|$ and its dimensionless derivative $a\, u_0'(r)/|u_m|$ are 
given by the solid and dotted lines in the upper panel of Figure 8, while 
the initial dimensionless tangential wind $v_0(r)/v_m$ and the initial 
dimensionless relative vorticity $a\, \zeta_0(r)/(4v_m)$ are given by the 
solid and dotted lines in the lower panel.  Plots of $u_0(r)$, $v_0(r)$, 
$w_0(r)$, and $\zeta_0(r)$ in dimensional form (obtained by using the radial
profiles shown in Figure 8 with the same
parameters as the single eyewall test case S5) are shown by the blue 
curves in Figure 10. This initial condition is very similar to the initial 
condition shown in Figure 1, except that there are secondary peaks (just 
outside $r=80$ km) in $u_0(r)$, $v_0(r)$, $w_0(r)$, and $\zeta_0(r)$. 
From Figure 8, note that there are two local minima in $u_0'(r)$, 
one at $r \approx 0.53\, a$ and one at $r \approx 1.36\, a$. Since $|u_0'(r)|$ 
is larger at the inner minimum, the inner shock will form before the outer 
shock.  The analytical solutions for this example, obtained by using these 
initial conditions in (\ref{eq4.5}), (\ref{eq4.7}), (\ref{eq4.8}), 
(\ref{eq4.15}), and (\ref{eq4.16}), are shown in Figures 9 and 10. 
As in the single eyewall case (test case S5), 
an inner shock develops at $r=17.6$ km and $t=1.69$ h. In addition, 
an outer shock has nearly developed by $t=1.69$ h. Because 
$u_0(\hat{r}_s)/u_0'(\hat{r}_s)$ is larger for the developing outer shock, 
the inward radial shift $(\hat{r}_s - r_s)$ predicted by (\ref{eq4.14}) is 
larger for the developing outer shock ($\hat{r}_s - r_s \approx 31$ km) 
than for the inner shock ($\hat{r}_s - r_s \approx 18$ km).  This model 
feature is consistent with the Hurricane Rita (2005) observations of 
\citet{houze07} and \citet{didlake11,didlake13} that show an outer 
eyewall with a larger outward tilt with height than the inner eyewall. It is 
also consistent with the structure produced in the model simulations of 
\citet{zhou09}. 
    
    One general conclusion that can be drawn from this simple solution is 
that double eyewalls naturally form when the radial profile of the boundary
layer inflow velocity is not monotonic outside the inner eyewall.  However,
one unrealistic aspect of the analytic model (\ref{eq4.1})--(\ref{eq4.2}) 
is that the radial equation of motion does not include the source term
proportional to $v-v_{\rm gr}$, so that the radial inflow simply damps along
characteristics.  In the more realistic numerical model results shown in
section 6, this source term is included so that in subgradient regions
(i.e., $v<v_{\rm gr}$) the radial inflow can increase along characteristics. 

    The analytical solutions found in this section contain singularities in the 
boundary layer pumping $w(r,t)$ and the vorticity $\zeta(r,t)$. Obviously, such 
singularities do not occur in nature; their mathematical existence reflects 
the simplicity of the physics included in (\ref{eq3.1})--(\ref{eq3.2}) 
and in (\ref{eq4.1})--(\ref{eq4.2}).   
In a nonhydrostatic, full-physics hurricane model, spikes in the radial 
distribution of boundary layer pumping might be expected to collapse to the
spatial scale of an individual cumulonimbus cloud, within which the vertical 
velocity would be limited by nonhydrostatic moist physics. 

     For the idealized analytical, double eyewall problem discussed here, 
the shock behavior is essentially determined by the nonzero initial condition 
$u_0(r)$.  In the next section we set $u_0(r)=0$ and allow the $u(r,t)$ 
field to develop through the $[f + (v+v_{\rm gr})/r](v - v_{\rm gr})$ term 
in (\ref{eq2.1}), with $v_{\rm gr}$ a specified function of $r$. 

%%%%%%%%%%%%%%%%%%%%%%%%%%%%%%%%%%%%%%%%%%%%%%%%%%%%%%%%%%%%%%%%%%%%%%%%%%%%%%%
\begin{figure}[!t]                          % Figure 11 (Model Forcing C1,C2,C3)
\centerline{\includegraphics[width=19pc]{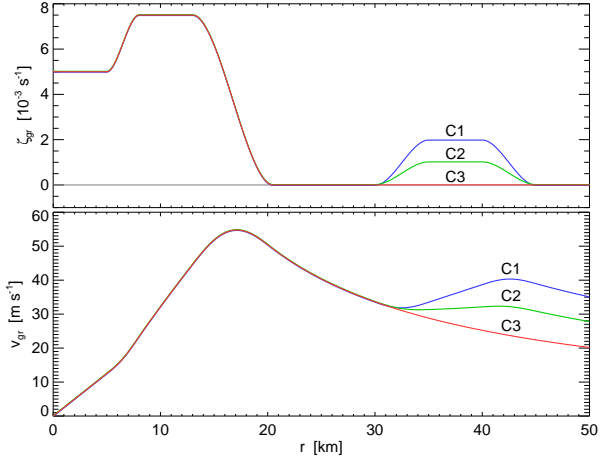}}
\caption{Radial distribution of the forcing $v_{\rm gr}(r)$ (bottom panel)
and the associated vorticity $\zeta_{\rm gr}(r)$ (top panel) for cases C1, C2,
and C3 of the numerical model. All three forcing profiles have the same
$\zeta_{\rm gr}(r)$ and the same $v_{\rm gr}(r)$ for $r \le 30$ km.}
\end{figure}

\begin{figure}[!b]                     % Figure 12 (Numerical Solution C1,C2,C3)
\centerline{\includegraphics[width=19pc]{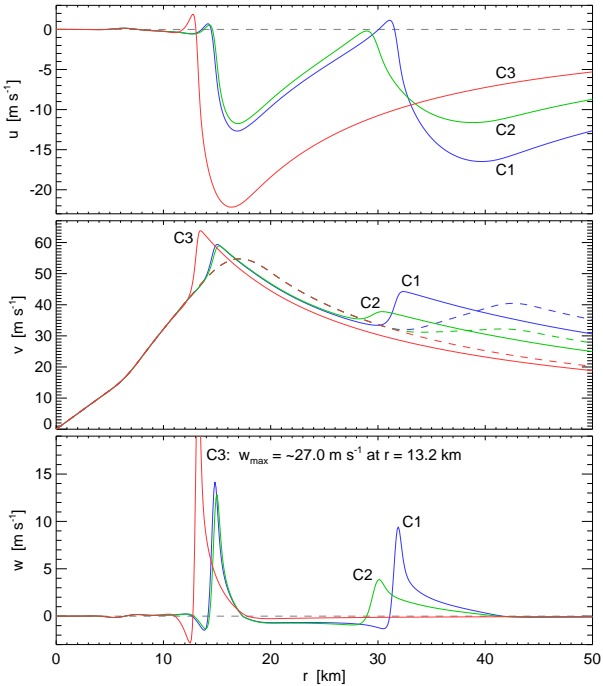}}
\caption{Steady state slab boundary layer radial profiles of radial velocity
$u$ (top panel), tangential velocity $v$ (middle panel), and vertical velocity
$w$ (bottom panel), for the three forcing profiles shown in Figure 11. The
radial profile of $w$ for the case with no concentric eyewall reaches a peak
of 27 m~s$^{-1}$, but has been cut off at 19 m~s$^{-1}$ for clarity of the
other profiles.}
\end{figure}
%%%%%%%%%%%%%%%%%%%%%%%%%%%%%%%%%%%%%%%%%%%%%%%%%%%%%%%%%%%%%%%%%%%%%%%%%%%%%%% 
\section{Numerical Solutions for Double Eyewalls}   %%%%% Section 6  %%%%%%

     In this section we present solutions of the problem
(\ref{eq2.1})--(\ref{eq2.7}), which has been solved numerically using 
centered, second-order spatial finite difference methods on the domain
$0 \le r \le 1000$ km with a uniform radial grid spacing of 100 m and a
fourth-order Runge-Kutta time differencing scheme with a time step of 1 s.  
The constants have been chosen as $h=1000$ m, $f=5.0\times 10^{-5}$ s$^{-1}$, 
and $K=1500$ m$^2$s$^{-1}$. The forcing has been designed to illustrate 
how the boundary layer flow transitions from one quasi-steady-state to another 
in response to an expansion of the balanced wind and vorticity fields above 
the boundary layer.

     Five numerical experiments have been performed. For the first three 
experiments, the forcing $v_{\rm gr}(r)$ is shown in the lower panel of Figure 11, 
with the associated $\zeta_{\rm gr}(r)$ shown in the upper panel. All three forcing 
profiles have the same $v_{\rm gr}(r)$ and the same $\zeta_{\rm gr}(r)$ for 
$r \le 30$ km. For experiments C1 and C2, the $\zeta_{\rm gr}(r)$ profiles 
have been locally ($30 < r < 45$ km) enhanced over that of experiment C3, 
so that the associated $v_{\rm gr}(r)$ profiles differ for $r > 30$ km. 
One can consider the sequence C3$\to$C2$\to$C1 as an enhancement of the 
outer gradient balanced flow while the inner core balanced flow remains 
unchanged. For each of these three specified $v_{\rm gr}(r)$ forcing 
functions, the numerical model was integrated until a steady state was 
obtained. Such steady states are generally obtained quickly, with most 
of the change from the initial conditions
$u(r,0)=0$ and $v(r,0)=v_{\rm gr}(r)$ occurring in the first hour, and only
small changes occurring after 3 hours. Figure 12 shows the steady state
boundary layer flows beneath each of these three forcing functions.  The three
panels show radial profiles ($0 \le r \le 50$ km) of the
boundary layer radial wind $u$ (top panel), tangential wind $v$ (middle panel),
and vertical velocity $w$ (bottom panel). Note that, for each case, strong
radial inflow, supergradient/subgradient tangential winds, and large boundary
layer pumping develop. Due to the $u(\partial u/\partial r)$ term in the
radial equation of motion, Burgers' shock-like structures develop just 
inside the local maxima in the initial tangential wind.  At the inner 
eyewall ($r \approx 16.5$ km) the maximum radial inflows are 22 m~s$^{-1}$ 
for case C3, 11.5 m~s$^{-1}$ for case C2, and 12.5 m~s$^{-1}$ for case C1,  
so the strength of the inner eyewall shock is considerably reduced by the
presence of an outer shock. Note that, even though cases C1 and C2 have 
stronger inflow than case C3 at $r \approx 40$ km, the situation is reversed 
at $r \approx 30$ km, a radius at which the radial inflow has been reduced 
to essentially zero for cases C1 and C2. Although the radial inflows for 
cases C1 and C2 do somewhat recover in the moat region between the two eyewalls 
($16.5 < r < 29$ km), the width of the moat and the strength of the forcing 
$[f + (v+v_{\rm gr})/r](v-v_{\rm gr})$ are not large enough to allow a full 
recovery of the radial inflow, leading to an inner eyewall boundary layer pumping
(bottom panel of Figure 12) that is reduced to approximately 50\% of the value 
obtained in case C3. 

%%%%%%%%%%%%%%%%%%%%%%%%%%%%%%%%%%%%%%%%%%%%%%%%%%%%%%%%%%%%%%%%%%%%%%%%%%%%%%%
\begin{figure}[!b]                          % Figure 13 (Model Forcing C3,C4,C5)
\centerline{\includegraphics[width=19pc]{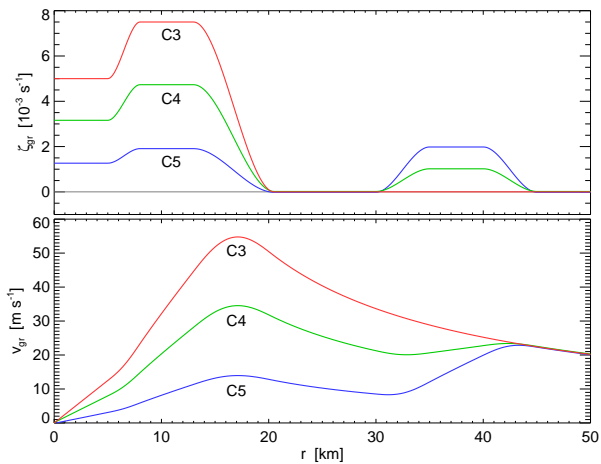}}
\caption{Radial distribution of the forcing $v_{\rm gr}(r)$ (bottom panel)
and the associated vorticity $\zeta_{\rm gr}(r)$ (top panel) for cases C3, C4,
and C5 of the numerical model. All three forcing profiles have the same
area-average vorticity inside $r=45$ km, so the corresponding $v_{\rm gr}(r)$
profiles are identical for $r \ge 45$ km.}
\end{figure}

\begin{figure}[!t]                     % Figure 14 (Numerical Solution C3,C4,C5)
\centerline{\includegraphics[width=19pc]{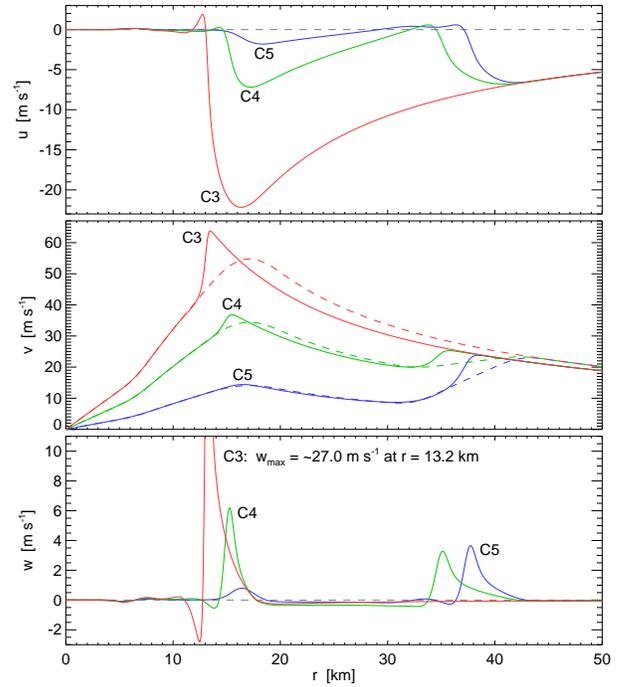}}
\caption{Steady state slab boundary layer radial profiles of radial velocity
$u$ (top panel), tangential velocity $v$ (middle panel), and vertical velocity
$w$ (bottom panel), for the three forcing profiles shown in Figure 13. The
radial profile of $w$ for the case with no concentric eyewall reaches a peak
of 27 m~s$^{-1}$, but has been cut off at 12 m~s$^{-1}$ for clarity of the
other profiles.}
\end{figure}
%%%%%%%%%%%%%%%%%%%%%%%%%%%%%%%%%%%%%%%%%%%%%%%%%%%%%%%%%%%%%%%%%%%%%%%%%%%%%%%
     The forcing functions for two additional experiments, denoted as C4 
and C5, are shown in Figure 13.  For reference, the previously discussed 
case C3 is also shown. Cases C3, C4, and C5 have the same $v_{\rm gr}(r)$ for 
$r > 45$ km. In the sequence C3$\to$C4$\to$C5, the inner eyewall vorticity 
decreases and the outer eyewall vorticity increases in such a way that the 
area averaged vorticity inside $r=45$ km remains unchanged. The results for 
cases C3, C4, and C5 are shown in Figure 14. Again, the radial inflow for 
cases C4 and C5 is reduced to essentially zero on the inside edge of the 
outer shock, and now the recovery of inflow in the moat is even weaker  
because the values of $v$ and $v_{\rm gr}$ are so nearly equal
that the $(v - v_{\rm gr})$ term in (\ref{eq2.1}) is unable to reestablish
significant inflow at the radius of the inner eyewall. In case C5, the result 
is a boundary layer pumping of less than 1 m~s$^{-1}$ at $r=16.5$ km. Thus, by
shutting off the radial inflow to the inner eyewall, the outer eyewall can
take over the role of the most diabatically active region. 
%Having encountered such flow structures in his three-dimensional  
%simulations, T. N. Krishnamurti has referred to this rapid reduction of the 
%outer eyewall boundary layer radial inflow as the ``Berlin wall effect."
 
    \citet{barnes83} argued that an outer spiral band or outer eyewall acts 
as a partial barrier, creating hostile conditions for the inner eyewall. 
The present results support this general line of reasoning, with added support 
for the idea that the partial barrier exists, not in the free troposphere, 
but in the frictional boundary layer. The hostile conditions for the inner 
eyewall result from a boundary layer shock-like feature produced at the outer eyewall. 
At the inner edge of this boundary layer feature, the radial inflow can be greatly
reduced, and, if the moat is narrow and the tangential velocity in the moat is 
only weakly subgradient, the radial inflow cannot sufficiently recover to 
provide for the maintenance of the inner eyewall. 
The results presented here also support the boundary layer control arguments  
of \citet{huang12} and \citet{abarca13}, which state that 
a radially expanding tangential wind field above the boundary layer is 
necessary for secondary eyewall formation and subsequent decay of the 
inner eyewall.

\section{Concluding Remarks}

     The results presented here provide some insight into questions such 
as: (1) What determines the size of the eye? (2) How are potential vorticity
rings produced?  (3) How does an outer concentric eyewall form and how does it
influence the inner eyewall? 
The slab boundary layer results support the notion that the size of the eye is 
determined by nonlinear processes that set the radius at which the eyewall
shock forms.  A boundary layer potential vorticity ring is also produced at
this radius. By boundary layer pumping and latent heat release, the boundary
layer PV ring is extended upward.  If, outside the eyewall, the boundary layer
radial inflow does not decrease monotonically with radius, a concentric eyewall
boundary layer shock can form. If it is strong enough and close enough to the
inner eyewall, this outer eyewall shock can choke off the boundary layer radial
inflow to the inner shock and effectively shut down the boundary layer pumping
at the inner eyewall. 

     The results presented here also emphasize the dual role played by
the surface stress terms. In the tangential equation of motion the
surface stress term decelerates the tangential flow, producing subgradient
flow ($v < v_{\rm gr}$).  In the radial equation of motion the
$(v - v_{\rm gr}) < 0$ term produces an inward radial flow down the pressure
gradient, which is favorable for shock formation. 
In contrast, the surface stress term in the radial equation of motion tends 
to retard shock formation. For storms of hurricane intensity and with a small 
enough radius of maximum gradient wind, this retarding effect is overcome by 
the shock generation process. 

     An issue that sometimes arises in the interpretation of modeling results
is whether hurricane development is a top-down or a bottom-up process
\citep{nolan07}.  One possible interpretation of the present results is that,
early in the development, the process is top-down as the free-tropospheric
vortex organizes and imposes a radial pressure gradient on the boundary layer.
Later, as a boundary layer shock forms, the process is bottom-up, with the
spike in boundary layer pumping setting the radius of the eyewall and hence
the size of the eye.

     Since the near-accident in Hurricane Hugo \citep{marks08}, research   
aircraft have not generally flown in the inner core boundary layer, which
means that this region of strong winds tends to be undersampled, although
the lack of flight level observations can be partially compensated through
the use of dropsondes and remote sensing from higher flight levels. 
An important challenge is the development of methods to safely obtain in-situ 
observations in the inner core boundary layer. 

     We have studied only the response to axisymmetric, nontranslating pressure 
fields with a constant depth, in which case the boundary layer shocks are circular. The problem of
the boundary layer response to a translating pressure field was pioneered by
\citet{chow71} and \citet{shapiro83}.  When the pressure field translates,
the shocks may become spiral shaped, as has been recently discussed by
\citet{williams12}. Also, as discussed by \citet{kepert10a, kepert10b}, the constant depth limits the slab boundary layer model's ability to
resolve important features of the tropical cyclone that are found in
height-resolving models. Another limitation of the present study is that we have
only explored the boundary layer dynamics of double eyewalls, whereas some
tropical cyclones can have more than two eyewalls. A well-documented case of
triple eyewalls was provided by \citet{mcnoldy04} for Hurricane Juliette (2001), 
which had peaks in relative vorticity at 9, 54, and 82 km.   
Similar dynamical concepts should apply to the understanding of triple
eyewalls, although the axisymmetry assumption is probably less valid for
the outermost eyewall. 

     In closing it is interesting to note that there are many hints of the 
existence of boundary layer shocks in the observational and modeling literature 
on hurricanes. One of the earliest comes from the insightful observational 
work of \citet{malkus58}. In describing the structure of the mature hurricane 
eye, she notes that ``the eye has several other mysterious features, the most
striking being the weak winds within it, despite the raging cyclonic gales in
the convective wall only a few kilometers away."  This description is
consistent with the U-shaped tangential wind profiles that evolve during
boundary layer shock formation.

%%% End of body of article:

%%%%%%%%%%%%%%%%%%%%%%%%%%%%%%%%
%% Optional Appendix goes here
%

\ifthenelse{\boolean{dc}}
{}
{\clearpage}
\begin{appendix}
\vspace{-12pt}
\section*{\begin{center}Characteristic Form\end{center}}

     Equations (\ref{eq2.1})--(\ref{eq2.7}) constitute a quasi-linear first
order system, i.e., the system is linear in the first derivatives but the
coefficients of these derivatives are functions of the dependent variables
$u$ and $v$. In the absence of the horizontal diffusion terms, these
equations constitute a hyperbolic system, which means that it can be rewritten
in characteristic form. Knowledge of the characteristic 
form allows for a deeper understanding of the way that characteristics can 
intersect and thereby produce discontinuities in $u$ and $v$ and singularities 
in $w$ and $\zeta$. To derive the characteristic form 
we shall rearrange (\ref{eq2.1}) and (\ref{eq2.2}) in such
a way that all the terms involving the derivatives $(\partial u/\partial t)$,
$(\partial u/\partial r)$, $(\partial v/\partial t)$, $(\partial v/\partial r)$
appear on the left-hand sides and all the other terms appear on the right-hand
sides. This procedure requires splitting the $w^-$ terms.
In regions where $w \ge 0$, the $w^-$ terms in (\ref{eq2.1}) and (\ref{eq2.2})
vanish. In regions where $w<0$, the $w^-$ terms do not vanish, in which case
these terms need to be expressed as $(\partial u/\partial r) + (u/r)$,
and then the $(\partial u/\partial r)$ parts need to be kept on to
the left-hand sides of (\ref{eq2.1}) and (\ref{eq2.2}) while the $(u/r)$ parts need
to be brought over to the right-hand sides. This procedure is
easily accomplished by noting that the mass continuity equation (\ref{eq2.3})
yields
\begin{equation}                                % Equation (A.1)
   w^- = \left(1-\alpha\right) h\left(\frac{\partial u}{\partial r} + \frac{u}{r}\right)
   \,\,\, {\rm where} \,\,\, 
          \alpha = \begin{cases}
                       1  & {\rm if} \,\,\, w \ge 0  \\
                       0  & {\rm if} \,\,\, w < 0,
                   \end{cases}
\label{eqA.1}
\end{equation}
which allows (\ref{eq2.1}) and (\ref{eq2.2}) to be written in the form
\begin{equation}                                % Equation (A.2)
        \frac{\partial u}{\partial t}
     + (2-\alpha)u\frac{\partial u}{\partial r} = F_1,
\label{eqA.2}
\end{equation}
\begin{equation}                                % Equation (A.3)
        \frac{\partial m}{\partial t} 
     + u\frac{\partial m}{\partial r}
     + \left(1-\alpha\right) (m -m_{\rm gr})\frac{\partial u}{\partial r} = F_2,
\label{eqA.3}
\end{equation}
where $m_\mathrm{gr}=rv_\mathrm{gr} + \frac{1}{2}fr^2$ is the gradient
absolute angular momentum and
\begin{equation}                                  % Equation (A.4)
    F_1 = -\frac{(1-\alpha) u^2}{r} 
        + \left(f + \frac{v+v_{\rm gr}}{r}\right) (v - v_{\rm gr})
        - c_D U\frac{u}{h},  
\label{eqA.4}
\end{equation}
\begin{equation}                                  % Equation (A.5)
    F_2 = - \frac{(1-\alpha) u(m - m_{\rm gr})}{r}
          - c_D U\frac{rv}{h},
\label{eqA.5}
\end{equation}
The forms (\ref{eqA.2}) and (\ref{eqA.3}) are convenient because the
nonlinearities associated with spatial derivatives are on the left-hand side
while all the other linear and nonlinear terms are on the right-hand side.
The classification of the system (\ref{eqA.2}) and (\ref{eqA.3}) as a
hyperbolic system and the determination of the characteristic form of this
system depends on finding the eigenvalues and left eigenvectors of the
matrix $A$, which is defined by
\begin{equation}                                  % Equation (A.6)
      A = \left(\begin{matrix}
                (2-\alpha)u           &   0   \\
                (1-\alpha)(m-m_{\rm gr})  &   u
                \end{matrix} \right)
\label{eqA.6}
\end{equation}
(see Chapter 5 of \cite{whitham74}).
Note that the matrix $A$ is composed of the coefficients of the
$(\partial u/\partial r)$ and $(\partial m/\partial r)$ terms on
the left-hand sides of (\ref{eqA.2}) and (\ref{eqA.3}).
For $n=1,2$, let $\left(\ell_1^{(n)} \,\, \ell_2^{(n)}\right)$ be the left
eigenvector of $A$ corresponding to the eigenvalue $\lambda^{(n)}$, i.e.,
\begin{equation}                                  % Equation (A.7)
   \left(\ell_1^{(n)} \,\,\, \ell_2^{(n)}\right)
          \left(\begin{matrix}
                (2-\alpha)u           &   0   \\
                (1-\alpha)(m-m_{\rm gr})  &   u
                \end{matrix} \right)
        = \lambda^{(n)}\left(\ell_1^{(n)} \,\,\, \ell_2^{(n)}\right).
\label{eqA.7}
\end{equation}
As is easily checked by direct substitution into (\ref{eqA.7}), the 
two eigenvalues and the two corresponding left eigenvectors are
\begin{equation}                                  % Equation (A.8)
  \begin{matrix}
   \lambda^{(1)} = (2-\alpha)u  & \Longleftrightarrow & \ell_1^{(1)} = 1,            \,\,\,  \ell_2^{(1)} = 0,  \\
   \lambda^{(2)} = u            & \Longleftrightarrow & \ell_1^{(2)} = m-m_{\rm gr}, \,\,\,  \ell_2^{(2)} = -u.
  \end{matrix}
\label{eqA.8}
\end{equation}
Since the eigenvalues $\lambda^{(1)}$ and $\lambda^{(2)}$ are real and the
corresponding left eigenvectors are linearly independent, the system
(\ref{eqA.2})--(\ref{eqA.3}) is hyperbolic and can be rewritten in
characteristic form. To obtain this characteristic form, we next
take the sum of $\ell_1^{(n)}$ times (\ref{eqA.2}) and $\ell_2^{(n)}$ times
(\ref{eqA.3}) to obtain
\begin{equation}                                % Equation (A.9)
  \begin{split}
     & \ell_1^{(n)}\left\{\frac{\partial u}{\partial t}
        + \left[(2-\alpha)u + (1-\alpha)(m-m_{\rm gr}) \frac{\ell_2^{(n)}}{\ell_1^{(n)}}\right]
                                       \frac{\partial u}{\partial r}\right\} \\
     &+ \ell_2^{(n)}\left\{\frac{\partial m}{\partial t} + u\frac{\partial m}{\partial r}\right\}
      = \ell_1^{(n)} F_1 + \ell_2^{(n)} F_2.
  \end{split}
\label{eqA.9}
\end{equation}
Using the eigenvector components given in (\ref{eqA.8}), equation (\ref{eqA.9})
becomes (for $n=1$ and $n=2$)
\begin{equation}                                % Equation (A.10)
                  \frac{\partial u}{\partial t}
     + (2-\alpha)u\frac{\partial u}{\partial r} = F_1,
\label{eqA.10}
\end{equation}
\begin{equation}                                % Equation (A.11)
       (m - m_{\rm gr})\left(\frac{\partial u}{\partial t}
                          + u\frac{\partial u}{\partial r}\right)
                  - u  \left(\frac{\partial m}{\partial t}
                          + u\frac{\partial m}{\partial r}\right) = F_3,
\label{eqA.11}
\end{equation}
where
\begin{equation}                                  % Equation (A.12)
  \begin{split}
    F_3 &= (m-m_{\rm gr}) F_1 - u F_2  \\
        &= \left(f + \frac{v+v_{\rm gr}}{r}\right)r\left(v - v_{\rm gr}\right)^2  
          + \frac{c_D U urv_{\rm gr}}{h}.
  \end{split}
\label{eqA.12}
\end{equation}
Since (\ref{eqA.10}) is identical to (\ref{eqA.2}), we conclude that
(\ref{eqA.2}) is already in characteristic form.
We now write (\ref{eqA.10}) and (\ref{eqA.11}) in the form
\begin{equation}                                % Equation (A.13)
        \frac{du}{dt} = F_1 \quad  {\rm on} \quad \frac{dr}{dt} = (2-\alpha)u,
\label{eqA.13}
\end{equation}
\begin{equation}                                % Equation (A.14)
        (m - m_{\rm gr})\frac{du}{dt} - u\frac{dm}{dt} = F_3
                           \quad  {\rm on} \quad \frac{dr}{dt} = u.
\label{eqA.14}
\end{equation}
Equations (\ref{eqA.13}) and (\ref{eqA.14}) constitute the characteristic 
form of the original system (\ref{eqA.2}) and (\ref{eqA.3}).
An advantage of (\ref{eqA.13}) and (\ref{eqA.14}) is that, along each family
of characteristic curves, the partial differential equations have been reduced
to ordinary differential equations. It is interesting to note that, in
regions of subsidence (i.e., where $\alpha=0$), information on $u$ is 
carried along characteristics given by $(dr/dt)=2u$, while information on 
a combination of $u$ and $m$ is carried along characteristics given by $(dr/dt)=u$. 
Thus, in regions of subsidence there are two distinct families of characteristics.
In contrast, for regions of boundary layer pumping (i.e., where $\alpha=1$), the two
families of characteristics become identical.

      Although in practice the forcing terms $F_1$ and $F_3$ are
too complicated to allow analytical solution of (\ref{eqA.13}) and
(\ref{eqA.14}), the numerical solution of these ordinary differential
equations can serve as the basis of the shock-capturing methods described 
by \citet{leveque02}.  In section 6 we have adopted the simpler approach of
solving (\ref{eq2.1})--(\ref{eq2.7}) using standard finite differences with
the inclusion of horizontal diffusion to control the solution near
shocks. Although this approach has some disadvantages
(e.g., unphysical oscillation near a shock), it provides a useful guide
to the expected results when full-physics hurricane models can be run
at the high horizontal resolution used here.

    In regions where $w<0$, we have $\alpha=0$ and the characteristic forms 
(\ref{eqA.13}) and (\ref{eqA.14}) distinguish two families of characteristics, 
one given by $(dr/dt)=2u$ and one given by $(dr/dt)=u$. In 
regions where $w \ge 0$, we have $\alpha=1$ and there is only one family 
of characteristics, given by $(dr/dt)=u$. In that case, (\ref{eqA.13}) 
can be used to eliminate $(du/dt)$ in (\ref{eqA.14}), which leads to the 
conclusion that $(du/dt)=F_1$ and $(dm/dt)=F_2$ on $(dr/dt)=u$. 
This case of only one family of characteristics is the one explored 
analytically in sections 3 and 4, with the forcing terms $F_1$ and $F_2$ 
set to zero in section 3, and with these forcing terms representing 
linear surface drag in section 4. 

    In passing we note that there is a less formal, more intuitive route
from (\ref{eqA.2}) and (\ref{eqA.3}) to the characteristic forms
(\ref{eqA.13}) and (\ref{eqA.14}). This intuitive route results from
simply noting that (\ref{eqA.2}) is already in characteristic form and
can be directly written as (\ref{eqA.13}), while the characteristic form
(\ref{eqA.14}) can be simply obtained by combining (\ref{eqA.2}) and
(\ref{eqA.3}) in such a way as to eliminate terms containing the factor
$(1-\alpha)(\partial u /\partial r)$.
\end{appendix}

%%%%%%%%%%%%%%%%%%%%%%%%%%%%%%%%%%%%%%%%%%%%%%%%%%%%%%%%%%%%%%%%
%
%  ACKNOWLEDGMENTS

\begin{acknowledgment}
     We would like to thank Joseph Biello, Paul Ciesielski, Mark DeMaria,
Alex Gonzalez, Hung-Chi Kuo, and Hugh Willoughby for their comments. This 
research has been supported by the Hurricane Forecast Improvement
Project (HFIP) through the Department of Commerce (DOC) National Oceanic and
Atmospheric Administration (NOAA) Grant NA090AR4320074 and through 
the National Science Foundation under Grants 
ATM-0837932 and AGS-1250966 and under the Science and Technology Center for Multi-Scale
Modeling of Atmospheric Processes, managed by Colorado State University through
cooperative agreement No.\ ATM-0425247.  The calculations were made on high-end
Linux workstations generously provided through a gift from the Hewlett-Packard
Corporation. This manuscript was generated using a modified version of the American
Meteorological Society \LaTeX\ template for journal page layout.
\end{acknowledgment}

% Create a bibliography directory and place your .bib file there.
% -REMOVE ALL DIRECTORY PATHS TO REFERENCE FILES BEFORE SUBMITTING TO THE AMS FOR PEER REVIEW
\ifthenelse{\boolean{dc}}
{}
{\clearpage}
\bibliographystyle{ametsoc2014}
\bibliography{2014arxiv_slocum}

%%%%%%%%%%%%%%%%%%%%%%%%%%%%%%%%%%%%%%%%%%%%%%%%%%%%%%%%%%%%%%%%%%%%%
% FIGURES-REMOVE ALL DIRECTORY PATHS TO FIGURE FILES BEFORE SUBMITTING TO THE AMS FOR PEER REVIEW
%%%%%%%%%%%%%%%%%%%%%%%%%%%%%%%%%%%%%%%%%%%%%%%%%%%%%%%%%%%%%%%%%%%%%

%
\end{document}